\documentclass[onecolumn]{IEEEtran} 
\usepackage{cite}

\ifCLASSINFOpdf
   \usepackage[pdftex]{graphicx}
   \graphicspath{{./figures/}}
   \DeclareGraphicsExtensions{.pdf,.jpeg,.png}
\else
   \usepackage[dvips]{graphicx}
   \graphicspath{{./figures/}}
   \DeclareGraphicsExtensions{.eps}
\fi
\usepackage{amsmath, bm, mathtools}
\newtheorem{lemma}{Lemma}

\newtheorem{definition}{Definition}

\newtheorem{example}{Example}
\newtheorem{remark}{Remark}
\newtheorem{illustration}{Illustration}

\usepackage{amssymb, textgreek}
\usepackage[subnum]{cases}

\DeclareMathOperator{\E}{\mathbb{E}}
\newcommand{\eqvclass}[3]{\mathrm{Cl}_{#1}^{(#3)}(#2)}
\newcommand{\classid}[2]{\mathrm{Cl}_{#1}^{(#2)}}

\makeatletter
\newcommand{\pushright}[1]{\ifmeasuring@#1\else\omit\hfill$\displaystyle#1$\fi\ignorespaces}
\newcommand{\pushleft}[1]{\ifmeasuring@#1\else\omit$\displaystyle#1$\hfill\fi\ignorespaces}
\makeatother

\usepackage{array}

\ifCLASSOPTIONcompsoc
  \usepackage[caption=false,font=normalsize,labelfont=sf,textfont=sf]{subfig}
\else
  \usepackage[caption=false,font=footnotesize]{subfig}
\fi

\usepackage{etoolbox}
\makeatletter
\AtBeginDocument{\patchcmd{\sf@subfloat}
  {\maincaptiontopfalse}
  {\maincaptiontoptrue}
  {}{}}
\makeatother

\usepackage{multirow}

\usepackage{url}

\begin{document}
%
\title{Zero-error Function Computation on a\\Directed Acyclic Network}
%
%
%

\author{Ardhendu~Tripathy,~\IEEEmembership{Student Member,~IEEE,}
        and~Aditya~Ramamoorthy,~\IEEEmembership{Member,~IEEE}
\thanks{This work was supported in part by the National Science Foundation (NSF) under grants CCF-1149860, CCF-1320416, CCF-1718470 and DMS-1120597. The material in this work has appeared in part at the 2016 IEEE International Symposium on Information Theory. The authors are with the Department of Electrical and Computer Engineering, Iowa State University, Ames, IA, 50011 USA. e-mails: \texttt{\{ardhendu,adityar\}@iastate.edu}}
}

\maketitle

\begin{abstract}
We study the rate region of variable-length source-network codes that are used to compute a function of messages observed over a network. The particular network considered here is the simplest instance of a directed acyclic graph (DAG) that is not a tree. Existing work on zero-error function computation in DAG networks provides bounds on the \textit{computation capacity}, which is a measure of the amount of communication required per edge in the worst case. This work focuses on the average case: an achievable rate tuple describes the expected amount of communication required on each edge, where the expectation is over the probability mass function of the source messages.

We describe a systematic procedure to obtain outer bounds to the rate region for computing an arbitrary demand function at the terminal. Our bounding technique works by lower bounding the entropy of the descriptions observed by the terminal conditioned on the function value and by utilizing the Schur-concave property of the entropy function. We evaluate these bounds for certain example demand functions. 
\end{abstract}


%
\IEEEpeerreviewmaketitle

\section{Introduction}
%
%
%
%
%
%
Computing functions of data observed over a network is a well-motivated problem, and different frameworks addressing the problem have been studied in the literature.
Broadly speaking, a general function computation problem can be modeled in the following manner. A directed acyclic graph (DAG) is used to model a communication network. Its vertices denote the nodes of the network that are assumed to have unrestricted computational power and storage capacity. The edges denote one-way communication links that can be thought of as noiseless \textit{bit-pipes}. Each node can act as a decoder on the information received on its incoming edges and as an encoder for information transmitted on its outgoing edges. Some of the nodes in the network, called the \textit{source} nodes, observe discrete-valued source \textit{messages} that take values in a finite alphabet. The random process generating the source messages is assumed to be stationary and memoryless. There is a single \textit{terminal} node that wishes to compute losslessly a discrete-valued function of all the source messages using the information it receives on its incoming edges. A solution to such a function computation problem will specify the communication carried out on each link and the information processing performed at each node of the network. We are interested in finding the minimal communication required for solving a given problem instance. The amount of communication in a network can be specified by a \textit{rate tuple} that has as many entries as the number of edges in the network. Each entry denotes the rate of the code employed on the corresponding edge in the network.
As remarked in \cite{giridhar-kumar}, this problem in its full generality encompasses several different areas in information theory, and as such, existing literature focuses on simplified versions that highlight different aspects of the problem.

If there is just one encoder and one decoder connected by a noiseless communication link, and the decoder wants to compute the identity function for the message, then it is a standard source coding problem. When there is some coded side information available through an additional link at the decoder, then the optimal rate pair is given by the Ahlswede--K\"orner--Wyner solution \cite[Thm.~10.2]{elGamal-kim-book}. Extending this one-help-one solution to a two-help-one scenario is not optimal, as was demonstrated for a particular two-help-one problem instance by K\"orner and Marton \cite{kornerM79}.

Consider now the case of two encoders, connected by two separate links to a decoder which is interested in computing the identity function on the pair of source messages. The optimal rate pair for this distributed source coding problem is known to be the Slepian--Wolf rate region, and this solution can be extended to multiple encoders each of which are directly connected to the decoder via a separate link \cite[Chap.~10]{elGamal-kim-book}.

In reproducing the source messages at the terminal in the above scenarios, the coding schemes allow for an $\epsilon$-block error that can be made as small as desired by choosing asymptotically large block lengths. Concurrently, work on zero-error source coding with side information was initiated by Witsenhausen \cite{witsenhausen76}. With $X$ being the source message and $Y$ the side information, he defined a \textit{confusability graph} $G_X$ for $X$ that specifies the realizations of $X$ that must be given distinct codewords by the encoder in order to attain zero-error. The optimum number of distinct codewords required was shown to be the graph chromatic number $\chi(G_X)$; for encoding multiple (say, $k$) instances it is $\chi(G_X^k)$ where $G_X^k$ denotes the $k$-wise AND product graph of $G_X$. Thus when $\chi(G_X^k) < (\chi(G_X))^k$, block encoding multiple instances allows one to reduce the number of distinct codewords required. The above idea along with the p.m.f. of $X$ was used to find the expected number of bits that must be transmitted by the encoder in \cite{alonO96}. For the case when the codewords are restricted to be prefix-free, their expected length was shown in \cite{alonO96} to be within one bit of the chromatic entropy $H_\chi(G_X, X)$ of the probabilistic graph $(G_X, X)$. When multiple instances are block encoded, the asymptotic per-instance expected codeword length is $\lim_{k \rightarrow \infty}\frac{1}{k}H_\chi(G_X^{\lor k}, X^k)$, where $G_X^{\lor k}$ is the $k$-wise OR product graph of $G_X$. This limit was shown to be the \textit{graph entropy} of $(G_X, X)$ in \cite{alonO96}. More information about the possible savings due to encoding multiple instances can be found in \cite[Sec.~XI]{kornerO98}.

In a function computing problem, the terminal would want to compute a general demand function of the messages and not necessarily the identity function. Reference \cite{orlitskyR01} considered computing a function on the setup of the source coding with side information problem, and allowed a vanishing block-error probability. They defined the conditional graph entropy $H(G_X, X|Y)$ for the probabilistic graph $(G_X, X|Y)$ of $X$ given $Y$ and showed that it is equal to the optimal number of bits per-instance of $X$ that must be communicated by the encoder when asymptotically large block lengths are used. This was extended by the authors in \cite{doshiSME10} where they defined a conditional chromatic entropy of a probabilistic graph and showed that $\lim_{k \rightarrow \infty}\frac{1}{k}H_\chi(G_X^{\lor k}, X^k | Y^k) = H(G_X, X|Y)$. This gave a graph coloring procedure to obtain codes with rate close to the lower bound of \cite{orlitskyR01}.

Function computation was also considered in the distributed setting where two encoders separately encode $X$ and $Y$ such that the decoder is able to compute $F(X, Y)$ losslessly. This scenario is closer to the source coding with side information problem than the distributed source coding problem because of the following. Any demand function can be computed at the decoder after communicating each of the source messages to it, thus the Slepian--Wolf region is an inner bound to the rate region of a function computation problem on the same network. K\"orner and Marton showed in \cite{kornerM79} that if the decoder wants to decode just $Z$, then the rate tuple $(R_X, R_Y, R_Z) = (H(Z), H(Z), 0)$ is achievable for the case when $Z = X \oplus Y$, where $X$ and $Y$ are a doubly symmetric binary source pair and $\oplus$ denotes modulo-$2$ sum. This rate tuple, with $R_Z = 0$, can be interpreted as the decoder wanting to compute the modulo-$2$ sum using the information obtained from the $X$ and $Y$ encoders. This view was taken by Han and Kobayashi in \cite{hanK87} where, subject to the constraint $\Pr(X=x,Y=y) > 0, \:\forall (x,y) \in \mathcal{X}\times \mathcal{Y}$, they gave necessary and sufficient conditions on the demand function $F(X,Y)$ for which the Slepian--Wolf region for the source pair $(X,Y)$ coincides with the rate region of the function computation problem. 

The authors in \cite{doshiSME10} also considered distributed function computation, where they showed that if the joint p.m.f $\Pr(X=x, Y=y)$ for $(x,y) \in \mathcal{X}\times \mathcal{Y}$ satisfied a restrictive `zigzag' condition, then coloring the confusability graphs $G_X^{\lor k}$ and $G_Y^{\lor k}$ and then using Slepian--Wolf code is optimal for asymptotic block length. This sufficient condition was relaxed in \cite{feiziM14} to a coloring connectivity condition by taking into account the function value. This condition was shown to characterize the rate region for function computation on one-stage tree-networks. It also gave an inner bound to the rate region of a general tree network. The rate region for a multi-stage tree-network when every source message at a vertex in the network satisfies a local Markovian property was characterized in \cite{STchamkerten16}. The function computation scenarios described above all allowed for a vanishing block-error probability with asymptotically large block lengths.

In this paper, we study the problem of zero-error function computation over a simple DAG network that is not a tree, shown in Figure \ref{fig:fig1}. We assume that the three source messages are independent and the terminal wants to compute an arbitrary specified demand function of the messages. We allow for block encoding and decoding of multiple instances at the sources and the terminal. Even with the independence assumption, characterizing the rate region of this function computation problem is difficult as was observed in \cite{tripathyR16}, where the problem of computing a particular \textit{arithmetic sum} demand function on the same DAG network was considered. 
We describe works closely related to our problem next.

\subsection{Related work}
Zero-error function computation over a graphical network using network coding \cite{nw-info-flow}, \cite{koetterM03} was studied in \cite{kowshikK12}. There they considered two variants of the communication load on the network, called \textit{worst-case} and \textit{average-case} complexity, depending on whether the probability information of the source messages was used or not. They characterized the rate region of achievable rate tuples that allowed zero-error function computation in tree-networks, each entry in a rate tuple was the rate of a code employed on the corresponding edge in the tree-network. They also made the observation that finding the rate region of a DAG network is significantly challenging because of multiple paths between a source node and the terminal, which allows for different ways of combining information at the intermediate nodes.

Zero-error function computation was also studied in \cite{appuswamyFKZ11}, where the authors defined the \textit{computation capacity} of a function computation problem instance. 
This is a generalization of the coding capacity of a network (\textit{c.f.} \cite[Sec.~VI]{cannonsDFZ06}, \cite{doughertyFZ06}), which is the supremum of the ratio $\frac{k}{n}$ over all achievable $(k,n)$ fractional coding solutions for that communication network. A $(k,n)$ fractional network code is one in which $k$ source messages are block encoded at each encoder and every edge in the network transmits $n$ symbols from the alphabet in one channel use. The authors in \cite{appuswamyFKZ11} characterized the computation capacity of multi-stage tree-networks by finding the necessary and sufficient amount of information that must be transmitted across all graph cuts that separate one or more source nodes from the terminal. Upper bounds on the computation capacity of DAG networks are more complicated and have been obtained in \cite{huangTY15}, \cite{guangYL16}. These upper bounds were shown to be unachievable for a function computation problem on a particular DAG in recent work \cite[Sec.~V]{guangYYL_arxiv17}. An illuminating example in the search for improved upper bounds for function computation on DAG networks has been the problem of computing the arithmetic sum of three source bits over the network shown in figure \ref{fig:fig1}. By counting the necessary and sufficient number of codewords required, the computation capacity for this problem was evaluated to be $\log_6 4$ in \cite[Sec.~V]{appuswamyFKZ11}. A qualitatively different line of work considers the computation of simple functions such as finite-field sum \cite{ramamoorthyL13,raiD12,tripathyR18} over arbitrary acyclic networks with multiple terminals, i.e., there is a set of terminals each of which is interested in recovering a function of certain source messages (see also \cite{doughertyFZ05,huangR12_TCOM,huangR14}) that discuss the multiple unicast problem which is an important special case of function computation). The literature in this area examines the role of the network topology in function computation.

While the upper bounds for DAG networks described above hold for the worst-case scenario, the arithmetic sum example (\textit{cf.} Example \ref{eg:fig1} in Section \ref{sec:formulation}) shows that we can do better in the average-case scenario by using the probability information of the source messages. It can be seen that an upper bound to the computation capacity corresponds to an outer bound to the rate region of a function computation problem. We use the framework of a \textit{source-network} code as used for zero-error network coding in \cite{songYC03}. Correspondingly, the quantity of interest here is the zero-error function computation rate region, and we provide outer bounds to this rate region for computing an arbitrary specified demand function on the network shown in figure \ref{fig:fig1}. We summarize our contributions below.
\subsection{Main contributions}
\begin{itemize}
\item We compute lower bounds on the rates that must be used on the edges of the DAG in Figure \ref{fig:fig1} in order to compute with zero-error an arbitrary demand function of the messages at the terminal. This provides us an outer bound to the achievable rate region.
\item The technique used for obtaining the lower bounds involves lower bounding the conditional entropy of the descriptions transmitted on the edges given the demand function value. This is done by first finding a family of p.m.fs. that must necessarily contain the conditional p.m.f. of the descriptions transmitted by any valid source-network code for the problem. The required lower bound was then obtained by finding the entropy-minimizing p.m.f. in this family. This p.m.f. was found using the Schur-concave property of the entropy function.
\item Computing the arithmetic sum over the DAG network in figure \ref{fig:fig1} was considered in \cite{tripathyR16}. By letting the number of symbols transmitted across the edges be a random variable $N$ (instead of a fixed $n$ for every $k$ source bits as in the $(k,n)$ fractional network code framework\footnote{As described in \cite{tripathyR16}, for finding the computation capacity, it can be assumed w.l.o.g. that the message vector $X_3^k$ is available to both the encoders at $s_1$ and $s_2$. Then the random variable $N$ is defined as a stopping time w.r.t. the pair of random variables transmitted over the edges $(s_1, t)$ and $(s_2, t)$.}), it was shown that the computation rate $\frac{k}{\mathbb{E} N} = 0.8$ is achievable. This is larger than its computation capacity in the $(k,n)$ fractional code framework, which was evaluated to be $\log_6 4$ in \cite{appuswamyFKZ11}. The work in \cite{tripathyR16} also gave an upper bound of $8/9$ for the computation capacity of this arithmetic sum example in the framework of \textit{variable-length} network codes. We recover and improve upon their results in the rate region framework, by obtaining a tighter outer bound to rate region in this case.
\item When the demand function to be computed over the DAG in figure \ref{fig:fig1} is set to be the $GF(2)$-sum, we show using a simple achievable scheme that the outer bound obtained for the rate region is tight.
\end{itemize}
The paper is organized as follows. 
Section \ref{sec:formulation} describes the problem setup formally and motivates the use of variable-length network codes as defined in Section \ref{sec:var_length_nw_code}. Section \ref{sec:outer_bd} describes the procedure used to obtain outer bounds to the rate region for computing an arbitrary demand function over the network in Figure \ref{fig:fig1}. Central to our approach are lower bounds to the conditional entropy of the descriptions transmitted, and these are described and illustrated using a running example in Section \ref{sec:lower_bound}. We also consider two other example demand functions in Sections \ref{subsec:arith_sum} and \ref{subsec:GF2_sum}. Section \ref{sec:conclude_future} concludes the chapter and lists some avenues for future work.

\section{Problem formulation}\label{sec:formulation}
\begin{figure}
\centering
\includegraphics[width=1.5in]{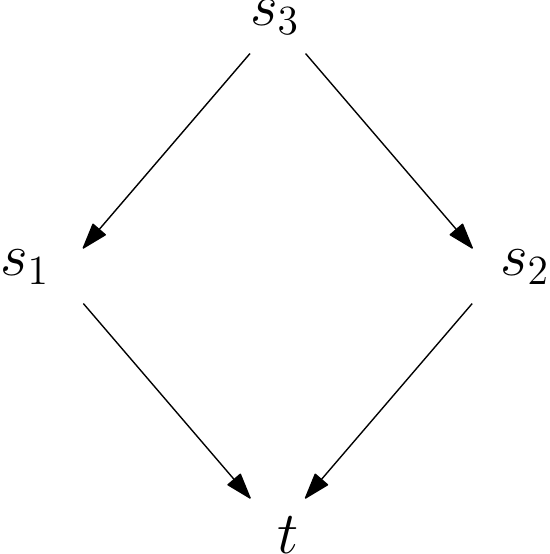}
\caption{A directed acyclic network with three sources, two of which also act as relay nodes, and one terminal. \label{fig:fig1}}
\end{figure}
The edges in Figure \ref{fig:fig1} (later denoted by an ordered pair of vertices) have unit-capacity, i.e., they transmit one symbol from $\mathcal{Z}$ in one time slot. We use the notation of a standard network code from \cite{appuswamyFKZ11}. In what follows, all logarithms denoted as $\log$ are to the base $2$ unless specified otherwise. Suppose that $\mathcal{Z}$ is the alphabet used for communication, and $|\mathcal{Z}|>1$. Vertices $s_1, s_2, s_3$ are the three memoryless source nodes that observe source messages $X_1,X_2,X_3$ respectively, each from a discrete alphabet $\mathcal{A}$ with size $|\mathcal{A}|>1$. The sources are assumed to be i.i.d. uniformly distributed over $\mathcal{A}$.
Terminal node $t$ wants to compute with zero error and zero distortion a known demand function:
\begin{IEEEeqnarray*}{RL}
f: \mathcal{A} \times \mathcal{A} \times \mathcal{A} &\rightarrow \mathcal{B},
\end{IEEEeqnarray*}
where $\mathcal{B}$ is a discrete alphabet of size $|\mathcal{B}|>1$. To avoid trivialities we assume that the demand function is not constant in any of its three arguments. A $(k,n)$ network code that satisfies the terminal has the following components.
\begin{itemize}
  \item An \textit{encoding function} $h^{e}(\cdot)$ associated with every edge $e$ in the network:
    \begin{IEEEeqnarray*}{RL}
    h^{(s_3,s_1)}(X_3^k):& \mathcal{A}^k \rightarrow \mathcal{Z}^n\\
    h^{(s_3,s_2)}(X_3^k):& \mathcal{A}^k \rightarrow \mathcal{Z}^n\\
    h^{(s_1,t)}\left(X_1^k, h^{(s_3,s_1)}(X_3^k)\right):& \mathcal{A}^k \times \mathcal{Z}^n \rightarrow \mathcal{Z}^n\\
    h^{(s_2,t)}\left(X_2^k, h^{(s_3,s_2)}(X_3^k)\right):& \mathcal{A}^k \times \mathcal{Z}^n \rightarrow \mathcal{Z}^n
    \end{IEEEeqnarray*}
    Here, the notation $X_j^k, j \in \{1,2,3\}$ denotes a block of $k$ i.i.d. uniform \textit{messages} observed at source node $s_j$.
  \item A \textit{decoding function} $\psi(\cdot)$ used by the terminal $t$:
  \begin{equation*}
    \psi\left(h^{(s_1,t)}(X_1^k, h^{(s_3,s_1)}(X_3^k)), h^{(s_2,t)}(X_2^k, h^{(s_3,s_2)}(X_3^k))\right): \mathcal{Z}^n \times \mathcal{Z}^n \rightarrow \mathcal{B}^k
  \end{equation*}
  With slight abuse of notation we let $f(X_1^k,X_2^k,X_3^k) \in \mathcal{B}^k$ denote the $k$ values returned by the demand function $f(X_1,X_2,X_3)$ when applied component-wise on a block of $k$ i.i.d. $(X_1,X_2,X_3)$ triples. That is,
  \begin{equation*}
      f(X_1^k,X_2^k,X_3^k) = \left(f(X_1^{(1)}, X_2^{(1)}, X_3^{(1)}), f(X_1^{(2)}, X_2^{(2)}, X_3^{(2)}), \ldots, f(X_1^{(k)}, X_2^{(k)}, X_3^{(k)})\right),
  \end{equation*} where $X_j^{(i)}$ denotes the $i$th component of $X_j^k$. By the zero-error criterion we have that
  \begin{equation*}
    \Pr\left\lbrace\psi\left(h^{(s_1,t)}(X_1^k, X_3^k), h^{(s_2,t)}(X_2^k, X_3^k)\right) \neq f(X_1^k,X_2^k,X_3^k)\right\rbrace = 0.
  \end{equation*}
\end{itemize}
If such a set of encoding and decoding functions exist for a choice of positive integers $k$ and $n$, we say that the network code \textit{computes} the demand function and has \textit{computation rate} $= k \log|\mathcal{A}|/(n\log |\mathcal{Z}|)$. The \textit{computation capacity} for a particular demand function is defined to be the supremum of all achievable computation rates.

The above framework is restrictive in the sense that the block length of the network code used (i.e., the value of $n$) is the same for each edge in the network. If we know the probability distribution associated with the message random variables and allow the block length of the network code on edge $e$ be a random variable $N_e$, then we can compress the descriptions transmitted on the edges and obtain savings in the expected length $\E [N_e]$.
\begin{example}\label{eg:fig1}
  Consider the problem of computing the sum over the integers (arithmetic sum) of three messages, i.e. $f(X_1, X_2, X_3) = X_1 + X_2 + X_3$, on the network shown in Figure \ref{fig:fig1}. The messages $X_1, X_2, X_3 \in \{0,1\}^k$ and $f(X_1, X_2, X_3) \in \{0,1,2,3\}^k$. This example was first considered in \cite{appuswamyFKZ11}, where they solve it using an optimal network code with $\mathcal{Z} = \{0,1\}$ that has computation rate $\log_6 4 \approx 0.77$. The optimal network code transmits the value of $X_3$ on the edges $(s_3, s_1), (s_3, s_2)$ and the values transmitted on the other two edges are as follows, assuming block length $k$ to be a multiple of $2$.
  \begin{align*}
    h^{(s_1,t)}(X_1, X_3) & = \begin{cases}
                                X_1^{(i)} + X_3^{(i)}, & \mbox{for all } 1\leq i \leq k/2 \\
                                X_1^{(i)}, & \mbox{otherwise}.
                              \end{cases} \\
    h^{(s_2,t)}(X_2, X_3) & =\begin{cases}
                              X_2^{(i)}, & \mbox{for all } 1\leq i \leq k/2 \\
                              X_2^{(i)}+X_3^{(i)}, & \mbox{otherwise}.
                            \end{cases}
  \end{align*}
  The number of bits to be transmitted on the edges using the above encoding functions can be seen to be $k(1 + \log 3)/2$. Now suppose that the messages are such that each $X_j^{(i)} \overset{\text{i.i.d.}}{\sim}$ Bern$(0.5)$, i.e., they are equally likely random bits. Then both $X_1^{(i)}+X_3^{(i)}$ and $X_2^{(i)}+X_3^{(i)}$ in the encoding functions above have a biased distribution and can be compressed. The entropy $H(X_1^{(i)}+X_3^{(i)}) = 1.5$, and hence its $\epsilon$-typical set \cite{cover-thomas} for $k/2$ instances can be enumerated using $1.5k/2$ bits. Thus the expected number of bits needed to be transmitted in this case can be seen to be
  \begin{equation*}
    \E N_{(s_1,t)} = \E N_{(s_2,t)} = (1 - \epsilon) (3k/4 + k/2) + \epsilon (k \log_2 3 + k/2) \approx 5k/4,
  \end{equation*}
  where the $\epsilon$ can be made small enough using large $k$. We note that $5/4 < (1 + \log 3)/2$, and that an analogous definition for the computation rate of a network code in the variable-length framework would give that $k/\max\{\E [N_{(s_1, t)}], \E [N_{(s_2,t)}]\} = 0.8 > 0.77$.
\end{example}
\begin{figure}
\centering
\includegraphics[scale=0.6]{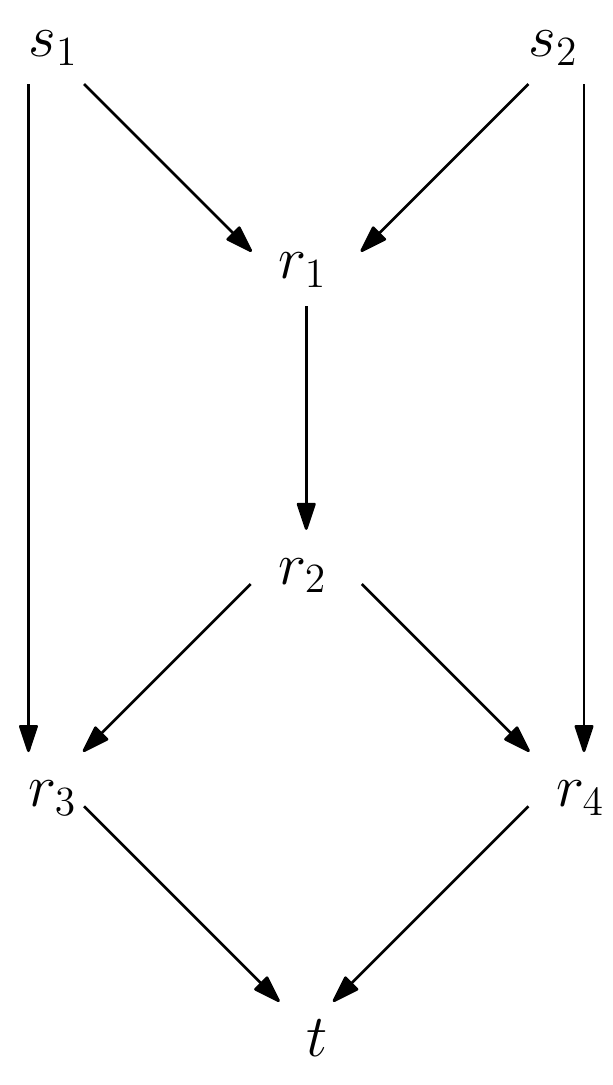}
\caption{A directed acyclic network with two sources, four relay nodes and one terminal. \label{fig:fig2}}
\end{figure}
\begin{example}\label{eg:fig2}
Consider computing the maximum over the integers $f(X_1, X_2) = \max \{X_1, X_2\}$, where $X_1, X_2 \in \{0, 1\}$, over the reverse butterfly network shown in Figure \ref{fig:fig2}. This problem was considered in \cite{guangYYL_arxiv17}, and they gave an upper bound of $2$ for the computation rate of any valid network code. They also gave a valid network code with rate $3/2$ that has the following edge functions, assuming block length $k$ to be a multiple of $3$.
\begin{align*}
h^{(s_1,r_3)}(X_1^k) &= (X_1^{(1)}, X_1^{(2)}, \ldots, X_1^{(k/3)}),\\
h^{(s_1,r_1)}(X_1^k) &= (X_1^{(1+k/3)}, X_1^{(2+k/3)}, \ldots, X_1^{(k)}),\\
h^{(s_2,r_4)}(X_2^k) &= (X_2^{(1)}, X_2^{(2)}, \ldots, X_2^{(k/3)}),\\
h^{(s_2,r_1)}(X_2^k) &= (X_2^{(1+k/3)}, X_2^{(2+k/3)}, \ldots, X_1^{(k)}),\\
h^{(r_1,r_2)}(h^{(s_1,r_1)}(X_1^k), h^{(s_2,r_1)}(X_2^k)) &= (\max\{X_1^{(1+k/3)}, X_2^{(1+k/3)}\}, \ldots, \max\{X_1^{(k)}, X_2^{(k)}\}),\\
h^{(r_2,r_3)}(h^{(r_1,r_2)}(h^{(s_1,r_1)}(X_1^k), h^{(s_2,r_1)}(X_2^k))) &= (\max\{X_1^{(1+k/3)}, X_2^{(1+k/3)}\}, \ldots, \max\{X_1^{(2k/3)}, X_2^{(2k/3)}\}),\\
h^{(r_2,r_4)}(h^{(r_1,r_2)}(h^{(s_1,r_1)}(X_1^k), h^{(s_2,r_1)}(X_2^k))) &= (\max\{X_1^{(1+2k/3)}, X_2^{(1+2k/3)}\}, \ldots, \max\{X_1^{(k)}, X_2^{(k)}\}),\\
h^{(r_3,t)}(h^{(s_1,r_3)}(\cdot), h^{(r_2,r_3)}(\cdot)) &= (h^{(s_1,r_3)}(\cdot), h^{(r_2,r_3)}(\cdot)),\\
h^{(r_4,t)}(h^{(s_2,r_4)}(\cdot), h^{(r_2,r_4)}(\cdot)) &= (h^{(s_2,r_4)}(\cdot), h^{(r_2,r_4)}(\cdot)),
\end{align*}
where the last two equations state that the edges $(r_3, t), (r_4,t)$ concatenate their inputs and forward it, and we have omitted the arguments of certain edge functions (represented by a dot) for brevity.

Now consider the case when each $X_j^{(i)} \overset{\text{i.i.d.}}{\sim}$ Bern$(0.75)$, i.e., it takes the value $1$ with probability $0.75$ and $0$ with probability $0.25$. Since $\max\{X_1^{(i)}, X_2^{(i)}\}$ has a biased p.m.f., we are able to compress the descriptions transmitted on the edges. We describe a variable-length network code that has a similar structure, except that the partitioning of the $k$ components of $X_1$ along the edges $(s_1, r_3)$ and $(s_1, r_1)$ is different from the $1$:$2$ ratio used before, and it uses typical set encoding. For the $i$th component, the entropy values $H(X_j^{(i)}) \approx 0.8113$ and $H(\max\{X_1^{(i)}, X_2^{(i)}\}) \approx 0.3373$ can be verified. Set the value $c \triangleq 1 / (2 - 0.3373 / (2\cdot 0.8113)) \approx 0.558$. Then the description transmitted on the edges $(s_1, r_3)$ and $(s_1,r_1)$ are as follows.
\begin{align*}
h^{(s_1,r_3)}(X_1) &= \mathcal{T}_{\epsilon}[(X_1^{(1)}, X_1^{(2)}, \ldots, X_1^{(k-ck)})]\\
h^{(s_1,r_1)}(X_1) &= \mathcal{T}_{\epsilon}[(X_1^{(1+k-ck)}, X_1^{(2+ck)}, \ldots, X_1^{(k)})],
\end{align*}
where $\mathcal{T}_{\epsilon}[\cdot]$ indicates typical set encoding and a similar description of $X_2$ is transmitted on the edges $(s_2, r_1)$ and $(s_2, r_4)$. This partitioning is chosen so as to have the best possible rate within a class of network codes that have the same structure as the initial network code described, as the following computations indicate.
\begin{align*}
\E N_{(s_1,r_1)} = \E N_{(s_2,r_1)} = 0.8113\cdot ck = 0.4527k,\: \E N_{(s_1, r_3)} = \E N_{(s_2, r_4)} = 0.8113 \cdot (1-c)k = 0.3586k,\\
\E N_{(r_1, r_2)} = 0.3373\cdot ck = 0.1882k,\: \E N_{(r_2, r_3)} = \E N_{(r_2, r_4)} = 0.1882k/2 = 0.0941k,\\
\E N_{(r_3, t)} = \E N_{(r_4, t)} = 0.0941k + 0.3586k = 0.4527k.
\end{align*}
Thus we have that $k/(\max_{\text{all edges} ~e} \E [N_e]) = k/(0.4527k) = 2.209 > 2$.
\end{example}
The above examples illustrate the reduction in communication load that can be achieved by using the knowledge of the probability distribution of the messages. The objective in this paper is to find bounds on rates of variable-length network codes that are valid for a given function computation problem. We first define the framework of variable-length network codes as adapted to the network in Figure \ref{fig:fig1} next.
\subsection{Variable-length network code for network in Figure \ref{fig:fig1}}\label{sec:var_length_nw_code}
We use the \textit{Source-Network Code} framework as described in \cite{songYC03} and adapt it to the function computation setting described above. The quantity of interest here is the rate region $\mathcal{R}$, which is a region containing all achievable rate tuples $\bm{R}$. Each rate tuple has four components, one for each edge in the network. We define the source-network code and the admissible rate tuples below.
\begin{definition}
Let $\mathcal{Z}^\ast$ denote the set of all finite-length sequences with alphabet $\mathcal{Z}$. A source-network code $\mathcal{C}_{f,k}$ for computing $f(X_1^k,X_2^k,X_3^k)$ in the network of Figure \ref{fig:fig1} has the following components:
\begin{enumerate}
    \item Encoding functions for edges $e \in \{(s_3, s_1), (s_3, s_2), (s_1,t), (s_2,t)\}$:
    \begin{align*}
    \phi_{(s_3,s_1)}(X_3^k)&: \mathcal{A}^k \rightarrow \mathcal{Z}^\ast\\
    \phi_{(s_3,s_2)}(X_3^k)&: \mathcal{A}^k \rightarrow \mathcal{Z}^\ast\\
    \phi_{(s_1,t)}\left(X_1^k, \phi_{(s_3,s_1)}(X_3^k)\right)&: \mathcal{A}^k \times \mathcal{Z}^{\ast} \rightarrow \mathcal{Z}^{\ast}\\
    \phi_{(s_2,t)}\left(X_2^k, \phi_{(s_3,s_2)}(X_3^k)\right)&: \mathcal{A}^k \times \mathcal{Z}^\ast \rightarrow \mathcal{Z}^\ast
\end{align*}
For brevity, we denote $\phi_{(s_1,t)}\left(X_1^k, \phi^{(s_3,s_1)}(X_3^k)\right)$ by the random variable $\bm{Z}_1$, and similarly define the r.v.s $\bm{Z}_2, \bm{Z}_{31}, \bm{Z}_{32}$.
    \item Decoding function for terminal $t$: $\psi_t:\mathcal{Z}^\ast \times \mathcal{Z}^\ast \rightarrow \mathcal{B}^k$ is such that $\Pr\{\psi_t(\bm{Z}_1, \bm{Z}_2) \neq f(X_1^k,X_2^k,X_3^k)\} =0$.
\end{enumerate}
\end{definition}
Thus the outputs of the encoders are variable length, and the terminal is equipped with a decoder that takes in a pair of variable length inputs and returns without any error the block of $k$ function computations on the message tuple. The rate of a source-network code is defined below, taking into account the different alphabets in which the messages and the codewords reside.
\begin{definition}
$\bm{R}=(R_{31}, R_{32}, R_1, R_2)$ is an admissible rate pair for the code $\mathcal{C}_{f,k}$ if for any $\epsilon > 0$ there exists a sufficiently large $k$ such that
\begin{align*}
\log |\mathcal{Z}|\E \ell(\bm{Z}_1) \leq k \log|\mathcal{A}|(R_1 + \epsilon)
\end{align*}
where $\E \ell(\bm{Z}_1)$ is the expected length (in symbols from $\mathcal{Z}$, and over the probability space of all message realizations) of the codeword $\bm{Z}_1$. A similar definition is used for the rates $R_{31}, R_{32}$ and $R_2$.
\end{definition}

\section{Bounds on the rate region for network in Figure \ref{fig:fig1}}\label{sec:outer_bd}
We use a lower bound on the expected length of the codewords transmitted on the edges in terms of their entropy. Theorem 3 in \cite{leung-yan-cheongC78} gives a lower bound for the expected length of a non-singular code over binary alphabet. The proof of the following lemma, in Appendix \ref{app:best-NS}, adapts their proof procedure for codes over non-binary alphabet $\mathcal{Z}$.
\begin{lemma}[Adapted from Theorem 3 in \cite{leung-yan-cheongC78}]\label{lem:best-NS}
The expected length (in symbols from an alphabet $\mathcal{Z}$) of the best non-singular code $\mathcal{C}_{\text{NS}}^{\star}(\bm{Z})$ for a random variable $\bm{Z}$ satisfies the following lower bound:
\begin{equation*}
\E \ell \left(\mathcal{C}_{\text{NS}}^{\star}(\bm{Z})\right) \geq H_{|\mathcal{Z}|}(\bm{Z}) - 2 \log_{|\mathcal{Z}|} \left(H_{|\mathcal{Z}| } (\bm{Z}) + |\mathcal{Z}|\right).
\end{equation*}
\end{lemma}
Since the identity mapping is also a non-singular code for $\bm{Z}$, we have that
\begin{equation}\label{eq:expect_length}
  \E \ell (\bm{Z}) = \sum_{\bm{z}} \Pr\{\bm{Z} = \bm{z}\}\ell(\bm{z}) \geq \E \ell \left( \mathcal{C}_{\text{NS}}^{\star}(\bm{Z}) \right).
\end{equation}
Because of the zero-error requirement for the decoding function at the terminal, we can give a value for what the sum rate $R_{31} + R_{32}$ must be greater than.
\begin{lemma}\label{lem:X_3eq}
Consider an equivalence relation on $\mathcal{A}^k$ for which $\bm{x}_3 \equiv \bm{x}_3'$ if and only if for all $(\bm{x}_1, \bm{x}_2) \in \mathcal{A}^k \times \mathcal{A}^k$, we have that $f(\bm{x}_1, \bm{x}_2, \bm{x}_3) = f(\bm{x}_1,\bm{x}_2,\bm{x}_3')$.
Define the function $g(X_3^k)$ which returns the equivalence class that $X_3^k$ belongs to under the above relation. Then the range of $g(X_3^k)$ is a subset of $\{1,2,\ldots, |\mathcal{A}|^k\}$ and we have that $R_{31} + R_{32} \geq H(g(X_3^k))/k\log|\mathcal{A}|$.
\end{lemma}
\begin{IEEEproof}
Suppose that $H_{|\mathcal{Z}|}(g(X_3^k)|\bm{Z}_{31},\bm{Z}_{32}) > 0$, then one cannot obtain $g(X_3^k)$ from the pair $(\bm{Z}_{31}, \bm{Z}_{32})$, i.e., there exist $\bm{x}_3 \not \equiv \bm{x}_3'$ but their associated codewords satisfy $\bm{z}_{31} = \bm{z}_{31}'$ and $\bm{z}_{32} = \bm{z}_{32}'$. There exists a pair $(\bm{x}_1, \bm{x}_2) \in \mathcal{A}^k \times \mathcal{A}^k$ such that $f(\bm{x}_1, \bm{x}_2, \bm{x}_3) \neq f(\bm{x}_1, \bm{x}_2, \bm{x}_3')$. However, since $\bm{z}_{31} = \bm{z}_{31}'$ and $\bm{z}_{32} = \bm{z}_{32}'$, the codewords transmitted on the edges $(s_1, t), (s_2, t)$ in the two cases satisfy $\bm{z}_1 = \bm{z}_1'$ and $\bm{z}_2 = \bm{z}_2'$. Thus the decoder receives the same input arguments in both the cases and consequently causes an error.

Thus we have that $H_{|\mathcal{Z}|}(g(X_3^k)|\bm{Z}_{31},\bm{Z}_{32}) = 0$. That gives us $H_{|\mathcal{Z}|}(g(X_3^k)) \leq H_{|\mathcal{Z}|}(\bm{Z}_{31}) +  H_{|\mathcal{Z}|}(\bm{Z}_{32})$, and using the upper bound to the entropy in terms of the expected codeword length (\textit{c.f.} equation \eqref{eq:expect_length} and lemma \ref{lem:best-NS}), we have the following.
\begin{align*}
  H_{|\mathcal{Z}|}(g(X_3^k)) \leq & \E \ell(\bm{Z}_{31}) + 2\log_{|\mathcal{Z}|}(H_{|\mathcal{Z}|}(\bm{Z}_{31}) + |\mathcal{Z}|) +\E \ell(\bm{Z}_{32}) + 2\log_{|\mathcal{Z}|}(H_{|\mathcal{Z}|}(\bm{Z}_{32}) + |\mathcal{Z}|),\\
  \implies H(g(X_3^k)) \leq & ~k(R_{31} + R_{32} + \epsilon)\log|\mathcal{A}|,
\end{align*} the second inequality uses the definition of rate, and $\epsilon$ can be made small enough because $H_{|\mathcal{Z}|}(\bm{Z}_{31})\leq H_{|\mathcal{Z}|}(X_3^k) = k$ and similarly for $H_{|\mathcal{Z}|}(\bm{Z}_{32})$.
\end{IEEEproof}
Accordingly, in the rest of the paper, we focus on the quantities $R_1$ and $R_2$.
Using inequality \eqref{eq:expect_length} and Lemma \ref{lem:best-NS}, we can conclude the following for the sum rate $R_1 + R_2$.
\begin{IEEEeqnarray}{RL}
\frac{\E \ell (\bm{Z}_1) + \E \ell(\bm{Z}_2)}{2} &\geq \frac{H_{|\mathcal{Z}|}(\bm{Z}_1) + H_{|\mathcal{Z}|}(\bm{Z}_2)}{2} - \log_{|\mathcal{Z}|}(H_{|\mathcal{Z}|}(\bm{Z}_1) + |\mathcal{Z}|) - \log_{|\mathcal{Z}|}(H_{|\mathcal{Z}|}(\bm{Z}_2) + |\mathcal{Z}|) \IEEEnonumber\\
&\overset{\text{(a)}}{\geq} \left(H_{|\mathcal{Z}|}(f(X_1^k,X_2^k,X_3^k)) + H_{|\mathcal{Z}|}(\bm{Z}_1, \bm{Z}_2|f(X_1^k,X_2^k,X_3^k))\right)/2 \IEEEnonumber\\
&\qquad - \log_{|\mathcal{Z}|}(H_{|\mathcal{Z}|}(\bm{Z}_1) + |\mathcal{Z}|) - \log_{|\mathcal{Z}|}(H_{|\mathcal{Z}|}(\bm{Z}_2) + |\mathcal{Z}|) \IEEEnonumber\\
&\overset{\text{(b)}}{\geq} \left(H_{|\mathcal{Z}|}(f(X_1^k,X_2^k,X_3^k)) + H_{|\mathcal{Z}|}(\bm{Z}_1, \bm{Z}_2|f(X_1^k,X_2^k,X_3^k), X_3^k)\right)/2 \IEEEnonumber\\
&\qquad - k\left(\log_{|\mathcal{Z}|} (H_{|\mathcal{Z}|}(\bm{Z}_1) + |\mathcal{Z}|) + \log_{|\mathcal{Z}|} (H_{|\mathcal{Z}|}(\bm{Z}_2) + |\mathcal{Z}|)\right)/k \IEEEnonumber\\
\implies \frac{R_1 + R_2 + \epsilon}{2} &\geq \frac{\E \ell (\bm{Z}_1) + \E \ell(\bm{Z}_2)}{2k\log_{|\mathcal{Z}|} |\mathcal{A}|} \IEEEnonumber\\
&\geq \frac{\alpha + H_{|\mathcal{Z}|}(f(X_1^k,X_2^k,X_3^k))/k}{2\log_{|\mathcal{Z}|} |\mathcal{A}|} - \epsilon' ~\text{for}~\epsilon, \epsilon' > 0~\text{and large}~k, \label{eq:sum-rate_lb}
\end{IEEEeqnarray}
where inequality (a) is because $H(\bm{Z}_1,\bm{Z}_2,f(X_1^k,X_2^k,X_3^k)) = H(\bm{Z}_1, \bm{Z}_2)$ by the zero error criterion, inequality (b) is true as conditioning reduces entropy, and implication \eqref{eq:sum-rate_lb} is true by the value of $\alpha$ obtained in Section \ref{sec:alpha}, by the definition of rate, and the fact that $H_{|\mathcal{Z}|}(\bm{Z}_1)\leq H(X_1^k,X_3^k) = 2k\log_{|\mathcal{Z}|}|\mathcal{A}|$ and similarly $H_{|\mathcal{Z}|}(\bm{Z}_2)\leq H(X_2^k,X_3^k) = 2k\log_{|\mathcal{Z}|}|\mathcal{A}|$.

Focusing on just $H_{|\mathcal{Z}|}(\bm{Z}_u)$ for either $u = 1$ or $2$, we have the following inequality.
\begin{IEEEeqnarray}{RL}
\mathbb{E}\ell(\bm{Z}_u) &\geq H_{|\mathcal{Z}|}(\bm{Z}_u) - 2\log_{|\mathcal{Z}|}(H_{|\mathcal{Z}|}(\bm{Z}_u) + |\mathcal{Z}|)\IEEEnonumber \\
&\overset{\text{(a)}}{\geq} H_{|\mathcal{Z}|}(\bm{Z}_u | X_3^k) - 2(\log_{|\mathcal{Z}|}(H_{|\mathcal{Z}|}(\bm{Z}_u)+|\mathcal{Z}|)) \overset{\text{(b)}}{\geq} \gamma k - 2k(\log_{|\mathcal{Z}|}(H_{|\mathcal{Z}|}(\bm{Z}_u)+|\mathcal{Z}|) / k)\IEEEnonumber \\
\implies R_u + \epsilon &\geq \log_{|\mathcal{A}|}|\mathcal{Z}|\mathbb{E}\ell(\bm{Z}_u)/k \geq \gamma/\log_{|\mathcal{Z}|}|\mathcal{A}| - \epsilon' ~\text{ for large } k, \IEEEyesnumber \label{eq:put-gamma}
\end{IEEEeqnarray} where inequality (a) is true because conditioning reduces entropy. A procedure to obtain the value of $\gamma$ in inequality (b) will be described in the next section (\textit{c.f.} equation \eqref{eq:gamma}).


\subsection{Lower bound on the conditional entropy} \label{sec:lower_bound}
In this section we use the structure of the demand function to obtain a lower bound on the conditional entropy of the descriptions transmitted on the edges $(s_3,s_1)$ and $(s_3,s_2)$. This enables us to find the values of $\alpha$ and $\gamma$ used in the previous section. To do this, we define quantities which denote the minimum number of distinct $(\bm{Z}_1, \bm{Z}_2)$-labels that allow the terminal to recover $f(X_1^k,X_2^k,X_3^k)$ with zero error. These quantities have been defined generally in  \cite{huangTY15}, \cite{guangYL16} and we adapt them to our particular network instance. Let lowercase letters $x_j, y_j \in \mathcal{A}$ denote  realizations of the message random variable $X_j, j \in \{1,2,3\}$. 
For ease of notation, the ordering of the arguments of $f(X_1,X_2,X_3)$ is ignored and subscripts indicate which message random variable a particular realization corresponds to.
\begin{definition}\label{def:partition_eq}
For two different message realizations $(x_1, x_2, x_3), (y_1, y_2, y_3) \in \mathcal{A}^3$ such that $x_3 = y_3 \triangleq a_3$, we say\footnote{Read as `$x_1$ is $a_3$-equivalent to $y_1$ for the message at $s_1$'.} that $x_1 \overset{a_3}{\equiv} y_1|_1$ if and only if $f(x_1,x_2,a_3)=f(y_1,y_2,a_3)$ for all $x_2 = y_2 \in \mathcal{A}$. Similarly, for two different message realizations $(x_1,x_2,x_3), (y_1,y_2,y_3) \in \mathcal{A}^3$ such that $x_3 = y_3 \triangleq a_3$, we say that $x_2 \overset{a_3}{\equiv} y_2|_2$ if and only if $f(x_1,x_2,a_3)=f(y_1,y_2,a_3)$ for all $x_1 = y_1 \in \mathcal{A}$.
\end{definition}
Note that for both $u=1,2$ and any value of $a_3 \triangleq x_3=y_3$,
\begin{itemize}
  \item $x_u=y_u$ implies $x_u \overset{a_3}{\equiv} y_u|_u$,
  \item $x_u \overset{a_3}{\equiv} y_u|_u$ implies $y_u \overset{a_3}{\equiv} x_u|_u$, and
  \item $x_u \overset{a_3}{\equiv} w_u|_u$ and $w_u \overset{a_3}{\equiv} y_u|_u$ implies $x_u \overset{a_3}{\equiv} y_u|_u$.
\end{itemize}
Thus $\overset{a_3}{\equiv}|_{u}$ is an equivalence relation on $\mathcal{A}$ for any demand function $f(X_1,X_2,X_3)$, choice of $u \in \{1,2\}$ and $a_3 \in \mathcal{A}$.
The number of equivalence classes of $\mathcal{A}$ induced by $\overset{a_3}{\equiv}|_{u}$ is denoted as $V_{u}(a_3)$ for both $u=1$ and $2$.
\begin{illustration}
Throughout the paper we will use a running example to illustrate the various quantities defined. In the network of Figure \ref{fig:fig1} we choose the message alphabet to be the finite field of order $3$ (denoted $GF(3)$) and the demand function\footnote{Any function from $GF(3)^3 \rightarrow GF(3)$ can be written as a multivariate polynomial in its arguments. Here it is $X_1^2X_2^2X_3 - X_1^2X_2X_3^2 + X_1X_2^2X_3^2 + X_1^2X_2X_3 + X_1X_2^2X_3 + X_1^2X_3^2 - X_2^2X_3^2 + X_1^2X_3 + X_2^2X_3 + X_1X_3^2 -X_2X_3^2 -X_1X_3 - X_2X_3 -X_3^2 + X_1 - X_2 + X_3$.} values are as listed in Table \ref{tab:illus}.
From Table \ref{tab:X3is1} we have that $0 \overset{1}{\equiv} 1|_1$ and $1 \overset{1}{\equiv} 2|_2$. 
This is because for any value of $X_2$, the entries in the rows corresponding to $X_1=0$ and $X_1=1$ in Table \ref{tab:X3is1} are the same, giving $0 \overset{1}{\equiv} 1|_1$. A similar statement is true for the columns corresponding to $X_2=1$ and $X_2=2$ in Table \ref{tab:X3is1}.
Since $0 \overset{1}{\equiv} 1 \overset{1}{\not \equiv} 2|_1$, we see that the relation $\overset{1}{\equiv}|_1$ partitions the alphabet $\mathcal{A}=GF(3)$ into two equivalence classes i.e. $\{0,1\}$ and $\{2\}$ and hence $V_1(1)=2$. From Table \ref{tab:illus} one can verify the following equivalence classes of $GF(3)$ for each kind of relation in the considered demand function.
\begin{IEEEeqnarray}{C}
\overset{0}{\equiv}|_1: \{0\} \cup \{1\} \cup \{2\},\:\: \overset{1}{\equiv}|_1: \{0,1\} \cup \{2\},\:\: \overset{2}{\equiv}|_1: \{0\} \cup \{1,2\}, \IEEEyesnumber \IEEEyessubnumber \label{eq:eq-part-index1}\\
\overset{0}{\equiv}|_2: \{0\} \cup \{1\} \cup \{2\},\:\: \overset{1}{\equiv}|_2: \{0\} \cup \{1,2\},\:\: \overset{2}{\equiv}|_2: \{0,1\} \cup \{2\}. \IEEEyessubnumber \label{eq:eq-part-index2}
\end{IEEEeqnarray}
Hence for this demand function we have that $V_1(0)=V_2(0)=3, V_1(1)=V_2(1)=2$ and $V_1(2)=V_2(2)=2.$
\end{illustration}
\begin{table*}[!t]
\centering
\caption{Function table for a demand function to be computed over the network in Figure \ref{fig:fig1}. The message alphabet is  $\mathcal{A}=GF(3)$. Table \ref{tab:X3is0} shows the function values for all $(X_1,X_2)$ pairs when $X_3=0$, table \ref{tab:X3is1} shows the function values when $X_3=1$ and table \ref{tab:X3is2} shows the function values when $X_3=2$.}
\label{tab:illus}
\subfloat[]{
       \begin{tabular}{|cc|ccc|}
         \hline
          \multicolumn{2}{|c|}{\multirow{2}{*}{$X_3=0$}}&   & $X_2$ &  \\
          \multicolumn{2}{|c|}{}&  $0$ & $1$ & $2$ \\ \hline
          & $0$ & $0$ & $2$ & $1$ \\
          $X_1$ & $1$ & $1$ & $0$ & $2$ \\
          & $2$ & $2$ & $1$ & $0$ \\
         \hline
       \end{tabular}
\label{tab:X3is0}}
\hfil
\subfloat[]{
       \begin{tabular}{|cc|ccc|}
         \hline
          \multicolumn{2}{|c|}{\multirow{2}{*}{$X_3=1$}}  &  & $X_2$ &  \\
          \multicolumn{2}{|c|}{}  & $0$ & $1$ & $2$ \\ \hline
          & $0$ & $0$ & $0$ & $0$ \\
          $X_1$ & $1$ & $0$ & $0$ & $0$ \\
          & $2$ & $1$ & $0$ & $0$ \\
         \hline
       \end{tabular}
\label{tab:X3is1}}
\hfil
\subfloat[]{
       \begin{tabular}{|cc|ccc|}
         \hline
          \multicolumn{2}{|c|}{\multirow{2}{*}{$X_3=2$}}  &  & $X_2$ &  \\
          \multicolumn{2}{|c|}{}  & $0$ & $1$ & $2$ \\ \hline
          & $0$ & $1$ & $1$ & $0$ \\
          $X_1$ & $1$ & $1$ & $1$ & $1$ \\
          & $2$ & $1$ & $1$ & $1$ \\
         \hline
       \end{tabular}
\label{tab:X3is2}}
\end{table*}

\begin{lemma}[Adapted from Lemma 3 in \cite{guangYL16}]\label{lem:partition-equiv}
The source node $s_1$ is the only source disconnected from the terminal if the edge $(s_1,t)$ is removed from the network, even though $s_3$ has a path connecting it to the terminal through the edge $(s_1,t)$. A similar statement also holds for the case of source node $s_2$ and edge $(s_2,t)$. For two realizations $(x_1, x_2,x_3), (y_1,y_2,y_3)\in \mathcal{A}^3$ such that $x_3=y_3=a_3$ a valid network code must transmit different $\bm{Z}_1$-labels on the edge $(s_1,t)$ if $x_1 \overset{a_3}{\not\equiv} y_1|_1$ and different $\bm{Z}_2$-labels labels on the edge $(s_2,t)$ if $x_2 \overset{a_3}{\not\equiv} y_2|_2$.
\end{lemma}
Thus, for a given realization $a_3$ of $X_3$, what matters is whether the function takes different values for different realizations of $X_1$ for some value of $X_2 \in \mathcal{A}$. If it does, then those two realizations of $X_1$ belong to different 
$\overset{a_3}{\equiv}|_1$ equivalence classes and hence by Lemma \ref{lem:partition-equiv} must have distinct labels transmitted over 
the edge $(s_1,t)$. A similar argument can be made for the labels transmitted on the $(s_2,t)$ edge based on the 
$\overset{a_3}{\equiv}|_2$ equivalence class of $X_2 \in \mathcal{A}$.
For either $u=1$ or $2$, the $\overset{a_3}{\equiv}|_u$ relation has a natural extension to vector realizations $\bm{x}_u \in \mathcal{A}^k$.
We then have the following lemma, whose proof is in Appendix \ref{app:proof_distinct}. We use lowercase boldface to denote vectors whose length is inferred from the context henceforth, like $\bm{a}_3 \triangleq \left(a_3^{(1)}, a_3^{(2)}, \ldots, a_3^{(k)}\right)$.
\begin{lemma}\label{lem:distinct_Z1_Z2_label}
Consider a block of $k$ independent realizations of $X_1,X_2$ and $X_3$ and let $\bm{a}_3 \in \mathcal{A}^k$ be the realization for $X_3^k$. Then for $u \in \{1,2\}$, the number of distinct $\bm{Z}_u$-labels that must be transmitted on the edge $(s_u,t)$ to allow the terminal to recover $f(X_1^k,X_2^k,X_3^k)$ with zero error is at least $V_u(\bm{a}_3) \triangleq \prod_{i=1}^{k}V_{u}(a_3^{(i)})$.
Equivalently, let $\overset{\bm{a}_3}{\equiv}|_u$ denote the collection of equivalence relations of Definition \ref{def:partition_eq} for each component of $\bm{a}_3$. In this notation, we have that
\begin{equation*}
\bm{x}_u \overset{\bm{a}_3}{\equiv} \bm{y}_u|_u \Leftrightarrow x_u^{(j)} \overset{a_3^{(j)}}{\equiv} y_u^{(j)}|_u, \quad \forall j \in \{1,2,\ldots, k\}.
\end{equation*}
If $\bm{x}_1 \overset{\bm{a}_3}{\not \equiv} \bm{y}_1|_1$, then $\phi_{(s_1,t)}(\bm{x}_1, \bm{a}_3) \neq \phi_{(s_1,t)}(\bm{y}_1, \bm{a}_3)$, i.e., their $\bm{Z}_1$ labels must be different. An analogous statement can be seen to be true for the $\bm{Z}_2$ label as well.
\end{lemma}

For either $u = 1$ or $2$ and each $i \in \{1,2,\ldots, k\}$, the equivalence classes under $\overset{a_3^{(i)}}{\equiv}|_u$ are denoted as $\eqvclass{u}{a_3^{(i)}}{j}$, where the superscript $j\in \{1,2,\ldots, V_u(a_3^{(i)})\}$ indexes the classes such that
\begin{equation*}
|\eqvclass{u}{a_3^{(i)}}{1}| \geq |\eqvclass{u}{a_3^{(i)}}{2}| \geq \cdots \geq |\eqvclass{u}{a_3^{(i)}}{V_u(a_3^{(i)})}|.
\end{equation*}
As described in the proof of Lemma \ref{lem:distinct_Z1_Z2_label}, $\mathcal{A}^k$ can be partitioned into $V_u(\bm{a}_3) = \prod_{i=1}^{k}V_u(a_3^{(i)})$ partitions based on the value of $\bm{a}_3$ for each component. Thus the partitions of $\mathcal{A}^k$ under $\overset{\bm{a}_3}{\equiv}|_u$ can be represented using a index vector $\bm{v}$ having $k$ components, each of which satisfies $v^{(i)}\in \{1, 2, \ldots, V_u(a_3^{(i)})\}$ and
\begin{equation*}
\bm{x}_u \in \eqvclass{u}{\bm{a}_3}{\bm{v}} \Leftrightarrow x_u^{(i)} \in \eqvclass{u}{a_3^{(i)}}{v^{(i)}}, \quad \forall i \in \{1,2, \ldots, k\}.
\end{equation*}
Similar to the scalar case, we add a subscript $t \in \{1, 2, \ldots, V_u(\bm{a}_3)\}$ to get the index vector $\bm{v}_t$ such that the equivalence classes under $\overset{\bm{a}_3}{\equiv}|_u$ satisfy
\begin{equation*}
|\eqvclass{u}{\bm{a}_3}{\bm{v}_1}| \geq |\eqvclass{u}{\bm{a}_3}{\bm{v}_2}| \geq \cdots \geq |\eqvclass{u}{\bm{a}_3}{\bm{v}_{V_u(\bm{a}_3)}}|.
\end{equation*}
From the definition, for every $t \in \{1, 2, \ldots, V_u(\bm{a}_3)\}$, we have that $\eqvclass{u}{\bm{a}_3}{\bm{v}_t} = \bigtimes_{i=1}^{k}\eqvclass{u}{a_3^{(i)}}{v_t^{(i)}}$, i.e., a cartesian product of $k$ scalar equivalence classes and accordingly $|\eqvclass{u}{\bm{a}_3}{\bm{v}_t}| = \prod_{i=1}^{k}|\eqvclass{u}{a_3^{(i)}}{v_t^{(i)}}|$.

The following notation is useful in characterizing all valid probability mass functions for $\bm{Z}_u$ for both $u=1,2$ and also the pair label $(\bm{Z}_1,\bm{Z}_2)$. For any index $i$ of a vector $\bm{p}$, we use $p^{[i]}$ to denote the $i$th component of $\bm{p}$ when it is arranged in non-increasing order.
For two vectors $\bm{p}, \bm{q}$ of the same length $l$, the vector $\bm{p}$ is majorized by $\bm{q}$, denoted as $\bm{p} \prec \bm{q}$, if the following holds.
\begin{IEEEeqnarray*}{Rl}
\sum_{i=1}^{t}p^{[i]} \leq \sum_{i=1}^{t}q^{[i]}, & ~\text{for all}~ t = 1,2,\ldots, l - 1, ~\text{and}\\
\sum_{i=1}^{l}p^{[i]} = \sum_{i=1}^{l}q^{[i]}. &
\end{IEEEeqnarray*} As an example, the vector $[0.5 ~~0.5]$ is majorized by $[0.25 ~~0.75]$. Note that any vector $\bm{p}$ is majorized by itself.

Lemma \ref{lem:distinct_Z1_Z2_label} gives a lower bound on the number of distinct $\bm{Z}_1$ and $\bm{Z}_2$ labels that must be used by a source-network code for a particular realization of $X_3^k$.
Using this and the uniform i.i.d. assumption on the messages, we can characterize the set of valid p.m.f.s for the conditional probability of the $\bm{Z}_u$ label given a realization $\bm{a}_3$ of the message $X_3^k$.
\begin{lemma}\label{lem:Z_u-majorize}
Let $\mathbb{Z}_{\geq 0}$ denote the set of non-negative integers and superscript $\top$ indicates the transpose operation. Consider the partitions of $\mathcal{A}^k$ induced by the set of relations $\overset{\bm{a}_3}{\equiv}|_u$ where $\bm{a}_3$ is a realization of $X_3^k$ and $u = 1$ or $2$. Let $L_u(\bm{a}_3) \geq V_u(\bm{a}_3)$ be the number of distinct $\bm{Z}_u$ labels assigned by an encoding scheme to the $|\mathcal{A}|^k$ different $(X_u^k, \bm{a}_3)$ pairs. Define a vector $\bm{d}_u(\bm{a}_3) \in \mathbb{Z}_{\geq 0}^{L_u(\bm{a}_3)}$ as
\begin{equation*}
\bm{d}_u(\bm{a}_3) \triangleq \begin{bmatrix}
                 |\eqvclass{u}{\bm{a}_3}{\bm{v}_1}| & |\eqvclass{u}{\bm{a}_3}{\bm{v}_2}| & \cdots & |\eqvclass{u}{\bm{a}_3}{\bm{v}_{V_u(\bm{a}_3)}}| & \bm{0}_{L_u(\bm{a}_3) - V_u(\bm{a}_3)}
               \end{bmatrix} ^\top,
\end{equation*} where $\bm{0}_{L_u(\bm{a}_3) - V_u(\bm{a}_3)}$ indicates a zero vector of length $L_u(\bm{a}_3) - V_u(\bm{a}_3)$. Then any valid conditional p.m.f. $\bm{p} \in \mathbb{R}^{L_u(\bm{a}_3)}$ for $\bm{Z}_u$ given the value of $X_3^k = \bm{a}_3$ satisfies $\bm{p} \prec \bm{d}_u(\bm{a}_3)/|\mathcal{A}|^k$.
\end{lemma}
\begin{IEEEproof}
We first note that $\bm{d}_u(\bm{a}_3)/|\mathcal{A}|^k$ is a valid p.m.f. as its components are non-negative and sum up to $1$. Suppose that there is an encoding scheme for $\bm{Z}_u$ such that $\Pr\{\bm{Z}_u | X_3^k = \bm{a}_3\} \triangleq \bm{p} \nprec \bm{d}_u(\bm{a}_3)/|\mathcal{A}|^k$. Furthermore let $\bm{p}$ be supported on $L_u(\bm{a}_3)$ components.
Then the assumption implies that there is a $t < L_u(\bm{a}_3)$ such that
\begin{equation*}
  \sum_{j=1}^{t} p^{[j]} > \frac{1}{|\mathcal{A}|^k}\sum_{j=1}^{t} d_u^{[j]} = \frac{1}{|\mathcal{A}|^k} \sum_{j=1}^{t}|\eqvclass{u}{\bm{a}_3}{\bm{v}_j}|.
\end{equation*}
Since each realization of $X_u^k$ is equally likely, the RHS in the above equation is the conditional probability given the value of $X_3^k = \bm{a}_3$ of the event that $X_u^k$ belongs to one of the $t$ largest equivalence classes under $\overset{\bm{a}_3}{\equiv}|_u$. The LHS is the conditional probability of observing any of the $t$ most probable $\bm{Z}_u$ labels. Hence the encoding scheme gives a total of $t$ distinct $\bm{Z}_u$ labels to at least as many $(X_u^k, \bm{a}_3)$ pairs for which $X_u^k$ belongs to $t+1$ different equivalence classes under $\overset{\bm{a}_3}{\equiv}|_u$. By the pigeonhole principle, this contradicts lemma \ref{lem:distinct_Z1_Z2_label}.
\end{IEEEproof}
In order to obtain a lower bound on $H(\bm{Z}_u|X_3^k = \bm{a}_3)$, we use the order-preserving property of the entropy function with respect to the majorization relation between two vectors \cite{MarshallOlkin_book}.
The entropy function $H: \mathbb{R}^{L_u(\bm{a}_3)} \rightarrow \mathbb{R}$ is a strictly Schur-concave function \cite[Chap. 3]{MarshallOlkin_book}, i.e., for two p.m.f.s  $\bm{p}, \bm{q} \in \mathbb{R}^{L_u(\bm{a}_3)}$ that are not equal to each other under any permutation of their components, we have that
\begin{equation*}
  \bm{p} \prec \bm{q} \implies H(\bm{p}) > H(\bm{q}).
\end{equation*}
Thus, from lemma \ref{lem:Z_u-majorize}, we have that $H_{|\mathcal{Z}|}(\bm{Z}_u|X_3^k = \bm{a}_3) \geq H_{|\mathcal{Z}|}(\bm{d}_u(\bm{a}_3)/|\mathcal{A}|^k)$. The value of $\gamma$ as used in equation \eqref{eq:put-gamma} can be found as follows.
\begin{equation}\label{eq:gamma}
  \gamma \triangleq \frac{1}{k}\sum_{\bm{a}_3 \in \mathcal{A}^k} \Pr(X_3^k = \bm{a}_3)H_{|\mathcal{Z}|}(\bm{d}_u(\bm{a}_3)/|\mathcal{A}|^k) \leq \frac{1}{k}\sum_{\bm{a}_3 \in \mathcal{A}^k}\Pr(X_3^k = \bm{a}_3)H_{|\mathcal{Z}|}(\bm{Z}_u|X_3^k = \bm{a}_3) = \frac{H_{|\mathcal{Z}|}(\bm{Z}_u|X_3^k)}{k}.
\end{equation}
\begin{illustration}
We evaluate the vector $\bm{d}_1(\bm{a}_3)$ for the example function and consequently obtain a lower bound for $H(\bm{Z}_1|X_3^k)$ and $R_1$ in this case. As shown in the previous illustration, $V_1(0)=3$ and $V_1(1) = V_1(2) = 2$. Suppose $X_3^k$ takes the value $\bm{a}_3$, where $\bm{a}_3$ has $m_t$ components with value $t$ for $t \in \{0,1,2\}$ such that $m_0 + m_1 + m_2 = k$.
For each component $a_3^{(i)} \in \mathcal{A}, i \in \{1, 2, \ldots, k\}$ the scalar equivalence classes under $\overset{a_3^{(i)}}{\equiv}|_u$ are given in equations \eqref{eq:eq-part-index1}, \eqref{eq:eq-part-index2}, and are represented as below.
\begin{align*}
&\eqvclass{u}{0}{1}=\{0\},\:\: \eqvclass{u}{0}{2}=\{1\},\:\: \eqvclass{u}{0}{3}=\{2\},\: \text{ for both } u=1,2 \text{ and}\\
&\eqvclass{1}{1}{1} = \{0,1\},\:\: \eqvclass{1}{1}{2} = \{2\},\:\: \eqvclass{1}{2}{1} = \{1,2\},\:\: \eqvclass{1}{2}{2} = \{0\},\\
&\eqvclass{2}{1}{1} = \{1,2\},\:\: \eqvclass{2}{1}{2} = \{0\},\:\: \eqvclass{2}{2}{1} = \{0,1\},\:\: \eqvclass{2}{2}{2} = \{2\}.
\end{align*}
Accordingly, the total number of partitions 
\begin{equation*}
  V_1(\bm{a}_3) = \textstyle \prod_{i=1}^{k} V_1(a_3^{(i)}) = \left(\textstyle \prod_{i:a_3^{(i)}=0}V_1(0)\right) \left(\textstyle \prod_{i:a_3^{(i)}=1}V_1(1)\right) \left(\textstyle \prod_{i:a_3^{(i)}=2}V_1(2)\right) = 3^{m_0}2^{m_1+m_2}.
\end{equation*}

To find the value of $|\eqvclass{1}{\bm{a}_3}{\bm{v}_1}|$, we pick the scalar partitions for each component that have the largest number of elements under $\overset{0}{\equiv}|_1, \overset{1}{\equiv}|_1$ and $\overset{2}{\equiv}|_1$.
From the above equations we get $|\eqvclass{1}{a_3^{(i)}}{v_1^{(i)}}|=2$ if $a_3^{(i)} \in \{1,2\}$ and $|\eqvclass{1}{a_3^{(i)}}{v_1^{(i)}}|=1$ if $a_3^{(i)}=0$. Thus $|\eqvclass{1}{\bm{a}_3}{\bm{v}_1}| = 2^{m_1 + m_2}$. Furthermore, since there are $3$ different largest equivalence classes under $\overset{0}{\equiv}|_u$, we get that
\begin{equation*}
  |\eqvclass{1}{\bm{a}_3}{\bm{v}_1}| = |\eqvclass{1}{\bm{a}_3}{\bm{v}_2}| = \cdots = |\eqvclass{1}{\bm{a}_3}{\bm{v}_{(3^{m_0})}}| = 2^{m_1+m_2}.
\end{equation*}
Now, we find the number of realizations $\bm{x}_1$ whose components do not necessarily belong to the largest scalar equivalence class at each component. If $t \leq m_1+m_2$ components of $\bm{x}_1$ are such that either $x_1^{(i)} = 0 \in \eqvclass{1}{a_3^{(i)}}{2}$ if $a_3^{(i)}=2$ or $\bm{x}_1^{(1)}=2 \in \eqvclass{1}{a_3^{(i)}}{2}$ if $a_3^{(i)}=1$, then the total number of $X_1^k$ realizations that belong to the same equivalence class as $\bm{x}_1$ is $2^{m_1+m_2-t}$. Thus we have that
\begin{IEEEeqnarray*}{sL}
for $t = 1$:& |\eqvclass{1}{\bm{a}_3}{\bm{v}_{3^{m_0}+1}}| = \cdots = |\eqvclass{1}{\bm{a}_3}{\bm{v}_{3^{m_0}+\binom{m_1+m_2}{1}3^{m_0}}}| = 2^{m_1+m_2-1},\\
for $t = 2$:& |\eqvclass{1}{\bm{a}_3}{\bm{v}_{3^{m_0}+\binom{m_1+m_2}{1}3^{m_0}+1}}| = \cdots = |\eqvclass{1}{\bm{a}_3}{\bm{v}_{3^{m_0}+\binom{m_1+m_2}{1}3^{m_0} + \binom{m_1+m_2}{2}3^{m_0}}}| = 2^{m_1+m_2-2},\\
\hphantom{for $t$}$\vdots$ \hphantom{$=2$}&\\
for $t = m_1+m_2$:& |\eqvclass{1}{\bm{a_3}}{\bm{v}_{3^{m_0}\sum_{l=0}^{m_1+m_2-1} \binom{m_1+m_2}{l}}}| = \cdots = |\eqvclass{1}{\bm{a}_3}{\bm{v}_{3^{m_0}\sum_{l=0}^{m_1+m_2}\binom{m_1+m_2}{l}}}| = 2^0 = 1.
\end{IEEEeqnarray*}
The above equations determine the components of the vector $\bm{d}_1(\bm{a}_3)$. We can then evaluate $H_{|\mathcal{Z}|}(\bm{d}_1(\bm{a}_3)/3^k)$ as
\begin{IEEEeqnarray*}{L}
3^{m_0}\cdot \frac{1}{3^{m_0}}\bigl(\frac{2}{3}\bigr)^{m_1+m_2}\log_{|\mathcal{Z}|} \frac{3^k}{2^{m_1+m_2}} + 3^{m_0}\binom{m_1+m_2}{1}\cdot \frac{1}{2\cdot 3^{m_0}}\bigl(\frac{2}{3}\bigr)^{m_1+m_2}\log_{|\mathcal{Z}|} \frac{3^k}{2^{m_1+m_2-1}}\\
+ 3^{m_0}\binom{m_1+m_2}{2}\cdot \frac{1}{2^2\cdot 3^{m_0}}\bigl(\frac{2}{3}\bigr)^{m_1+m_2}\log_{|\mathcal{Z}|} \frac{3^k}{2^{m_1+m_2-2}} + \cdots + 3^{m_0}\binom{m_1+m_2}{m_1+m_2}\cdot \frac{1}{2^{m_1+m_2}\cdot 3^{m_0}}\bigl(\frac{2}{3}\bigr)^{m_1+m_2}\log_{|\mathcal{Z}|} \frac{3^k}{2^{0}}\\
= (k\log_{|\mathcal{Z}|} 3-(m_1+m_2)\log_{|\mathcal{Z}|}2)\biggl(\frac{2}{3}\biggr)^{m_1+m_2}\biggl(1 + \binom{m_1+m_2}{1}2^{-1} + \cdots + \binom{m_1+m_2}{m_1+m_2}2^{-m_1-m_2}\biggr)\\
+ \log_{|\mathcal{Z}|}2 \biggl(\frac{2}{3}\biggr)^{m_1+m_2}\biggl(\binom{m_1+m_2}{1}2^{-1} + 2\binom{m_1+m_2}{2}2^{-2} + \cdots + (m_1+m_2)\binom{m_1+m_2}{m_1+m_2}2^{-m_1-m_2}\biggr)\\
= (k\log_{|\mathcal{Z}|} 3 - (m_1 + m_2)\log_{|\mathcal{Z}|}2)\biggl(\frac{2}{3}\biggr)^{m_1+m_2}\biggl(\frac{3}{2}\biggr)^{m_1+m_2} + \log_{|\mathcal{Z}|}2 \biggl(\frac{2}{3}\biggr)^{m_1+m_2} \frac{m_1+m_2}{3}\biggl(\frac{3}{2}\biggr)^{m_1+m_2} \\
= k\log_{|\mathcal{Z}|} 3 - \frac{2(m_1+m_2)}{3}\log_{|\mathcal{Z}|}2.
\end{IEEEeqnarray*}
It follows that
\begin{IEEEeqnarray*}{RL}
H_{|\mathcal{Z}|}(\bm{Z}_1|X_3^k) &= \sum_{\bm{a}_3}\Pr\{X_3^k = \bm{a}_3\}H_{|\mathcal{Z}|}(\bm{Z}_1|X_3^k = \bm{a}_3)
=\!\! \sum_{\substack{m_1,m_2 = 0,\\m_1+m_2 \leq k}}^{k}~\sum_{\substack{\bm{a}_3 \:\text{has}\:m_1\:\text{1's},\\m_2\:\text{2's}}}\Pr\{X_3^k = \bm{a}_3\}H_{|\mathcal{Z}|}(\bm{Z}_1|X_3^k = \bm{a}_3)\\
&\geq \sum_{\substack{m_1, m_2 = 0,\\m_1+m_2 \leq k}}^{k}\frac{k!}{3^k m_1!m_2!(k-m_1-m_2)!}\biggl(k\log_{|\mathcal{Z}|} 3 - \frac{2(m_1+m_2)}{3}\log_{|\mathcal{Z}|}2 \biggr)\\
&= k\log_{|\mathcal{Z}|}3 - \frac{2\log_{|\mathcal{Z}|}2}{3}\frac{2k3^{k-1}}{3^k} = k(\log_{|\mathcal{Z}|} 3 - \frac{4}{9}\log_{|\mathcal{Z}|}2).
\end{IEEEeqnarray*}
Thus the value of $\gamma$ is $(\log_{|\mathcal{Z}|} 3 - \frac{4}{9}\log_{|\mathcal{Z}|}2)$ and from equation \eqref{eq:put-gamma}, $R_1 + \epsilon \geq \log_{|\mathcal{A}|}3 - \frac{4}{9}\log_{|\mathcal{A}|}2 \approx 0.7196.$
\end{illustration}

To obtain a lower bound on the conditional entropy of the pair of labels as in equation \eqref{eq:sum-rate_lb}, we characterize the family of valid conditional p.m.f.s for the  pair $(\bm{Z}_1, \bm{Z}_2)$  given the values of the demand function $f(X_1^k,X_2^k,X_3^k)$ and the realization of the message $X_3^k$. 
Towards this end we first find the number of distinct $(\bm{Z}_1,\bm{Z}_2)$-labels that must be assigned by the network code to message tuples that result in a particular value, say, $\bm{b} \in \mathcal{B}^k$ of the demand function $f(X_1^k,X_2^k,X_3^k)$. The set $A_3(\bm{b})$ has all possible realizations $\bm{a}_3$ of $X_3^k$ that can result in the value of $\bm{b}$ for the demand function, i.e.,
\begin{equation*}
  A_3(\bm{b}) \triangleq \{\bm{a}_3\in \mathcal{A}^k: \exists\: \bm{x}_1,\bm{x}_2 \in \mathcal{A}^k \text{ such that } f(\bm{x}_1,\bm{x}_2,\bm{a}_3)=\bm{b}\}.
\end{equation*}

Let $M(\bm{a}_3, \bm{b})$ denote the number of distinct $(\bm{Z}_1,\bm{Z}_2)$ pair labels used for message tuples that have $X_3^k = \bm{a}_3$ and $f(\bm{x}_1,\bm{x}_2,\bm{a}_3)=\bm{b}$.
Consider two message tuples $(\bm{x}_1,\bm{x}_2,\bm{a}_3)$ and $(\bm{y}_1,\bm{y}_2,\bm{a}_3)$ which satisfy $f(\bm{x}_1, \bm{x}_2, \bm{a}_3)=f(\bm{y}_1, \bm{y}_2, \bm{a}_3)=\bm{b}$.
If either $\bm{x}_1 \overset{\bm{a}_3}{\not\equiv} \bm{y}_1|_{1}$ or $\bm{x}_2 \overset{\bm{a}_3}{\not\equiv} \bm{y}_2|_{2}$, then the pair of labels $(\bm{Z}_1, \bm{Z}_2)$ assigned to the two message tuples must be different.
This motivates us to define the pair index set:
\begin{equation*}
  \mathcal{V}_{12}(\bm{a}_3, \bm{b}) \triangleq
  \left\lbrace (\classid{1}{\bm{v}_j}, \classid{2}{\bm{w}_t}) :
  \parbox{0.53\textwidth}{\centering
   $\exists \: \bm{x}_1 \in \eqvclass{1}{\bm{a}_3}{\bm{v}_j}, \bm{x}_2 \in \eqvclass{2}{\bm{a}_3}{\bm{w}_t} \text{ s.t. } f(\bm{x}_1, \bm{x}_2, \bm{a}_3)=\bm{b},$\\ $1\leq j \leq V_1(\bm{a}_3), 1\leq t \leq V_2(\bm{a}_3)$}
 \right\rbrace.
\end{equation*}
The above discussion then implies that $M(\bm{a}_3, \bm{b}) \geq |\mathcal{V}_{12}(\bm{a}_3, \bm{b})|$.
By the definition, if $(\classid{1}{\bm{v}}, \classid{2}{\bm{w}}) \in \mathcal{V}_{12}(\bm{a}_3, \bm{b})$ then for every $i \in \{1,2, \ldots, k\}$, the pair of scalar equivalence classes $(\classid{1}{v^{(i)}}, \classid{2}{w^{(i)}}) \in \mathcal{V}_{12}(a_3^{(i)}, b^{(i)})$.

\begin{illustration}
Consider block size $k=1$ and a realization $b=0$ of the demand function of Table \ref{tab:illus}. In the previous illustration we evaluated that $\eqvclass{1}{1}{1}=\{0,1\}$, $\eqvclass{1}{1}{2}=\{2\}$ and $\eqvclass{2}{1}{1}=\{1,2\}$, $\eqvclass{2}{1}{2}=\{0\}$.
Then we can evaluate that the pair index set $\mathcal{V}_{12}(1, 0)=\{(\classid{1}{1},\classid{2}{1}), (\classid{1}{1},\classid{2}{2}),(\classid{1}{2},\classid{2}{1})\}$. Note that the pair $(\classid{1}{2},\classid{2}{2}) \notin \mathcal{V}_{12}(1, 0)$ as the elements of that pair of equivalence classes do not result in the demand function value of $0$, i.e.,
\begin{equation*}
  \eqvclass{1}{1}{2}=\{2\},\: \eqvclass{2}{1}{2}=\{0\}\:\text{ but for }\: x_1 = 2, x_2 = 0, x_3 = 1,\: f(x_1,x_2,x_3)=1 \neq 0.
\end{equation*}
Thus in this case we have that $|\mathcal{V}_{12}(1,0)| =3$. The other pair index sets are summarized in Table \ref{tab:eq-class_coupling}.
\begin{table}
  \centering
  \setlength{\tabcolsep}{4pt}
  \caption{The sets $\mathcal{V}_{12}(a_3, b)$ for different values of $a_3$ and $b$ in different rows and columns respectively.}\label{tab:eq-class_coupling}
  \begin{tabular}{c|ccc}
    \hline
    $a_3$ & $b=0$ & $b=1$ & $b=2$ \\\hline
    $0$ & $\{(\classid{1}{1},\classid{2}{1}),(\classid{1}{2},\classid{2}{2}),(\classid{1}{3},\classid{2}{3})\}$ & $\{(\classid{1}{1},\classid{2}{3}),(\classid{1}{2},\classid{2}{1}),(\classid{1}{3},\classid{2}{2})\}$ & $\{(\classid{1}{1},\classid{2}{2}),(\classid{1}{2},\classid{2}{3}),(\classid{1}{3},\classid{2}{1})\}$ \\
    $1$ & $\{(\classid{1}{1},\classid{2}{1}),(\classid{1}{1},\classid{2}{2}),(\classid{1}{2},\classid{2}{1})\}$ & $\{(\classid{1}{2},\classid{2}{2})\}$ & $\emptyset$ \\
    $2$ & $\{(\classid{1}{2},\classid{2}{2})\}$ & $\{(\classid{1}{1},\classid{1}{1}),(\classid{1}{1},\classid{2}{2}),(\classid{1}{2},\classid{2}{1})\}$ & $\emptyset$ \\
    \hline
  \end{tabular}
\end{table}
\end{illustration}

We now explicitly derive a p.m.f. whose entropy is a lower bound to the conditional entropy $H(\bm{Z}_1, \bm{Z}_2 | f(X_1^k, X_2^k, X_3^k) = \bm{b}, X_3^k = \bm{a}_3)$. Let $A_{123}(\bm{b}) \subseteq \mathcal{A}^k \times \mathcal{A}^k \times \mathcal{A}^k$ contain all message tuples that are present in the pre-image of the demand function value of $\bm{b}$. We also let $A_{123}(\bm{b}, \bm{a}_3) \subseteq A_{123}(\bm{b})$ denote the subset of message tuples for which the realization of $X_3^k$ is $\bm{a}_3$.
Suppose that $(\classid{1}{\bm{v}},\classid{2}{\bm{w}}) \in \mathcal{V}_{12}(\bm{a}_{3},\bm{b})$. The number of different message tuples that cause the membership of the equivalence class pair $(\classid{1}{\bm{v}},\classid{2}{\bm{w}}) \in \mathcal{V}_{12}(\bm{a}_{3},\bm{b})$ is denoted as follows.
\begin{IEEEeqnarray}{rL}
h_{\bm{a}_{3}}(\bm{v},\bm{w}) &\triangleq \left|\{(\bm{x}_1,\bm{x}_2): \bm{x}_1 \in \eqvclass{1}{\bm{a}_{3}}{\bm{v}}, \bm{x}_2 \in \eqvclass{2}{\bm{a}_{3}}{\bm{w}}, f(\bm{x}_1,\bm{x}_2,\bm{a}_{3})=\bm{b}\}\right|, \label{eq:h_a-defn}\\
&= |\eqvclass{1}{\bm{a}_{3}}{\bm{v}}|\cdot |\eqvclass{2}{\bm{a}_{3}}{\bm{w}}|, \label{eq:h_a-value}\\
&= \prod_{i=1}^{k} |\eqvclass{1}{a_3^{(i)}}{v^{(i)}}|\cdot |\eqvclass{2}{a_3^{(i)}}{w^{(i)}}| = \prod_{i=1}^{k} h_{a_3^{(i)}}(v^{(i)}, w^{(i)}). \label{eq:num-msg-tuples_vecv-vecw}
\end{IEEEeqnarray}
Equality \eqref{eq:h_a-value} is true above as by Definition \ref{def:partition_eq} every element of an equivalence class under $\overset{\bm{a}_{3}}{\equiv}|_1$ results in the same demand function value (while the other message $\bm{x}_2$ is held constant), and since $(\classid{1}{\bm{v}},\classid{2}{\bm{w}}) \in \mathcal{V}_{12}(\bm{a}_{3}, \bm{b})$, there is at least one $\bm{x}_1 \in \eqvclass{1}{\bm{a}_{3}}{\bm{v}}$ and one $\bm{x}_2 \in \eqvclass{2}{\bm{a}_{3}}{\bm{w}}$ such that $f(\bm{x}_1,\allowbreak \bm{x}_2,\allowbreak \bm{a}_{3})=\bm{b}$. Hence every other pair of elements in $\eqvclass{1}{\bm{a}_{3}}{\bm{v}} \times \eqvclass{2}{\bm{a}_{3}}{\bm{w}}$ would also result in the same demand function value with $X_3^k = \bm{a}_{3}$.

\begin{illustration}
For block size $k=1$ and demand function realization $b=0$, we can check that $A_3(0)=\{0,1,2\}$. Following the indexing of the equivalence partitions and Table \ref{tab:eq-class_coupling} we have that 
$h_1(1,1) = |\eqvclass{1}{1}{1}|\cdot |\eqvclass{2}{1}{1}|=4$ as
$\eqvclass{1}{1}{1} = \{0,1\}$ and $\eqvclass{2}{1}{1} = \{1,2\}$.
One can similarly check that $h_1(1,2)=h_1(2,1)=2$ and for other values of $a_3$, that 
$h_0(1,1)=h_0(2,2)=h_0(3,3)=1$ and $h_2(2,2)=1$.

For block size $k = 3$ and $\bm{b}=(1,2,1)$ we can check that $\bm{a}_3 = (0,0,1) \in A_3(\bm{b})$.
Then the equivalence class pairs under $\bm{a}_3$ that result in this $\bm{b}$ can be obtained in the following manner. If $(\classid{1}{\bm{v}}, \classid{2}{\bm{w}}) \in \mathcal{V}_{12}(\bm{a}_3, \bm{b})$, then for each component, using Table \ref{tab:eq-class_coupling} we have that
\begin{itemize}
\item $(\classid{1}{v^{(1)}}, \classid{2}{w^{(1)}}) \in \mathcal{V}_{12}(a_3^{(1)}, b^{(1)}) = \mathcal{V}_{12}(0,1) = \{(\classid{1}{1},\classid{2}{3}),(\classid{1}{2},\classid{2}{1}),(\classid{1}{3},\classid{2}{2})\}$,
\item $(\classid{1}{v^{(2)}}, \classid{2}{w^{(2)}}) \in \mathcal{V}_{12}(a_3^{(2)}, b^{(2)}) = \mathcal{V}_{12}(0,2) = \{(\classid{1}{1},\classid{2}{2}),(\classid{1}{2},\classid{2}{3}),(\classid{1}{3},\classid{2}{1})\}$,
\item $(\classid{1}{v^{(3)}}, \classid{2}{w^{(3)}}) \in \mathcal{V}_{12}(a_3^{(3)}, b^{(3)}) = \mathcal{V}_{12}(1,1) = \{(\classid{1}{2},\classid{2}{2})\}$.
\end{itemize}
As $(\classid{1}{\bm{v}}, \classid{2}{\bm{w}}) = (\bigtimes_{i=1}^{3} \classid{1}{v^{(i)}}, \bigtimes_{i=1}^{3} \classid{2}{w^{(i)}})$, we get that $|\mathcal{V}_{12}(\bm{a}_3, \bm{b})| = 9$. One of these nine equivalence class pairs is the pair $(\classid{1}{(1,1,2)}, \classid{2}{(3,2,2)})$.
Then using equation \eqref{eq:num-msg-tuples_vecv-vecw} we have that
\begin{equation*}
h_{(0,0,1)}((1,1,2),(3,2,2)) = h_0(1,3) \cdot h_0(1,2) \cdot h_1(2,2) = 1\cdot 1\cdot 1 = 1.
\end{equation*}
\end{illustration}
The proof of the following lemma is similar in spirit to that of Lemma \ref{lem:Z_u-majorize}.
\begin{lemma}
For any $\bm{a}_3 \in A_3(\bm{b})$, define the vector
\begin{equation*}
\bm{h}_{\bm{b}, \bm{a}_3} \triangleq \begin{bmatrix}
h_{\bm{a}_{3}}\left((\bm{v},\bm{w})_1\right) & h_{\bm{a}_{3}}\left((\bm{v},\bm{w})_2\right) & \cdots & h_{\bm{a}_{3}}\left((\bm{v},\bm{w})_{|\mathcal{V}_{12}(\bm{a}_{3}, \bm{b})|}\right) & \bm{0}_{M(\bm{a}_3, \bm{b})-|\mathcal{V}_{12}(\bm{a}_{3}, \bm{b})|}
\end{bmatrix}^\top,
\end{equation*} where $\bm{0}_t$ indicates a vector of zeros of length $t$  and subscript $j$ in $(\bm{v},\bm{w})_j$ indexes all the equivalence class pairs such that
\begin{equation*}
h_{\bm{a}_{3}}\left((\bm{v},\bm{w})_1\right) \geq h_{\bm{a}_{3}}\left((\bm{v},\bm{w})_2\right) \geq \cdots \geq  h_{\bm{a}_{3}}\left((\bm{v},\bm{w})_{|\mathcal{V}_{12}(\bm{a}_{3}, \bm{b})|}\right).
\end{equation*}
Then all conditional probability mass functions $\bm{p} \in \mathbb{R}^{M(\bm{a}_3,\bm{b})}_{\geq 0}$ on $M(\bm{a}_3, \bm{b})$ valid $(\bm{z}_1,\bm{z}_2)$-labels given the value $\bm{b}$ of the demand function and the realization $\bm{a}_3$ of $X_3^k$ satisfy $\bm{p} \prec \bm{h}_{\bm{b}, \bm{a}_3}/|A_{123}(\bm{b}, \bm{a}_3)|$.
\end{lemma}
It will be useful for analysis to define the following index sets of a demand function realization $\bm{b}\in \mathcal{B}^k$ and the message realization $\bm{a}_3 \in A_3(\bm{b})$.
\begin{definition}\label{def:index_sets}
For every $p \in \mathcal{B}$, let $\mathcal{I}_p(\bm{b}) \subseteq \{1,2,\ldots, k\}$ be the index set of components of the demand function realization $\bm{b}$ that are equal to $p$, i.e.,
\begin{equation*}
  \mathcal{I}_p(\bm{b}) = \{i : b^{(i)} = p, i \in \{1,2,\ldots, k\}\}.
\end{equation*}
 We can similarly define the index set $\mathcal{J}_q(\bm{x}_3)$ for every $q \in \mathcal{A}$.
\end{definition}
\begin{illustration}
We evaluate $H(\bm{h}_{\bm{b}, \bm{a}_3}/|A_{123}(\bm{b}, \bm{a}_3)|)$ for a general block length $k$ of the example demand function. To specify $\bm{h}_{\bm{b}, \bm{a}_3}$, we need to find $|\mathcal{V}_{12}(\bm{a}_3, \bm{b})|$ and $h_{\bm{a}_3}((\bm{v}, \bm{w})_j)$ for all $j \in \{1,2,\ldots, |\mathcal{V}_{12}(\bm{a}_3, \bm{b})|\}$.
Suppose the realization $\bm{b} \in \{0,1,2\}^k$ has $m_1$ 1's, $m_2$ 2's and $k-m_1-m_2$ 0's, and the realization of $X_3^k$ is some $\bm{a}_3 \in A_3(\bm{b})$.
Let $t_{p,q}$ for any $p \in \mathcal{B}, q \in \mathcal{A}$ be defined as $t_{p,q}\triangleq |\mathcal{I}_p(\bm{b}) \cap \mathcal{J}_q(\bm{a}_3)|$. Note that $t_{2,1}=t_{2,2}=0$ for any choice of $\bm{b}$ and $\bm{a}_3$.

\paragraph{$|\mathcal{V}_{12}(\bm{a}_3, \bm{b})|$}
We find the number of non-zero components of $\bm{h}_{\bm{b}, \bm{a}_3}$, i.e. $|\mathcal{V}_{12}(\bm{b}, \bm{a}_3)|$, as follows.
For every $(\classid{1}{\bm{v}}, \classid{2}{\bm{w}}) \in \mathcal{V}_{12}(\bm{a}_3, \bm{b})$, we have from Table \ref{tab:eq-class_coupling} that for the index $i$, 
\begin{itemize}
  \item if $i \in (\mathcal{I}_0(\bm{b}) \cap \mathcal{J}_2(\bm{a}_3)) \cup (\mathcal{I}_1(\bm{b}) \cap \mathcal{J}_1(\bm{a}_3))$, there is only one possible choice for the scalar equivalence class pair $(\classid{1}{v^{(i)}}, \classid{2}{w^{(i)}})$,
  \item else if $i$ is not in the previous index set, then there are three possible choices for the scalar equivalence class pair $(\classid{1}{v^{(i)}}, \classid{2}{w^{(i)}})$.
\end{itemize}
Thus the total number of equivalence class pairs for for the choice of $\bm{b}$ and $\bm{a}_3$ is $|\mathcal{V}_{12}(\bm{a}_3, \bm{b})| = 3^{k - t_{0,2} - t_{1,1}}$. Next we evaluate the components of the vector $\bm{h}_{\bm{b}, \bm{a}_3}$.

\paragraph{$h_{\bm{a}_3}((\bm{v}, \bm{w})_1)$}
Based on the subscript indexing of the pair $(\bm{v}, \bm{w})$ and equation \eqref{eq:num-msg-tuples_vecv-vecw}, the value of $h_{\bm{a}_3}((\bm{v}, \bm{w})_1)$ is obtained by counting the number of message tuples in the equivalence class pair that has the largest scalar equivalence class pair for each component. As evaluated in the previous illustration, the largest equivalence class pair for every $i \in \mathcal{I}_0(\bm{b}) \cap \mathcal{J}_1(\bm{a}_3)$ has $h_1(1,1)=4$ message tuples, and for every $i \in \mathcal{I}_1(\bm{b}) \cap \mathcal{J}_2(\bm{a}_3)$ the largest equivalence class again has $h_2(1,1)=4$ message tuples. For all other values of $i$, the largest equivalence class has a single input tuple. Hence we have that $h_{\bm{a}_3}((\bm{v}, \bm{w})_1) = 4^{t_{0,1}+t_{1,2}}$. Note that there are three different choices for the largest scalar equivalence class when $i$ belongs to one of the following sets.
\begin{itemize}
\item $i \in \mathcal{I}_0(\bm{b})\cap \mathcal{J}_0(\bm{a}_3)$: there are $t_{0,0} = k-m_1-m_2-t_{0,1}-t_{0,2}$ such components,
\item $i \in \mathcal{I}_1(\bm{b})\cap \mathcal{J}_0(\bm{a}_3)$: there are $t_{1,0} = m_1-t_{1,1}-t_{1,2}$ such components, and
\item $i \in \mathcal{I}_2(\bm{b})\cap \mathcal{J}_0(\bm{a}_3)$: there are $t_{2,0} = m_2$ such components.
\end{itemize}
Thus there are $3^{k-m_1-m_2 - t_{0,1} - t_{0,2}} \cdot 3^{m_1 - t_{1,1} - t_{1,2}} \cdot 3^{m_2} = 3^{k - t_{0,1} - t_{0,2} - t_{1,1} - t_{1,2}}$ components of $h_{\bm{b}, \bm{a}_3}$ which have the same value. Let $k' \triangleq k - t_{0,1} - t_{0,2} - t_{1,1} - t_{1,2}$. Hence we have that
\begin{equation*}
h_{\bm{a}_3}((\bm{v}, \bm{w})_1) = h_{\bm{a}_3}((\bm{v}, \bm{w})_2) = \cdots = h_{\bm{a}_3}((\bm{v}, \bm{w})_{3^{k'}}) =  4^{t_{0,1}+t_{1,2}}.
\end{equation*}

\paragraph{$h_{\bm{a}_3}((\bm{v}, \bm{w})_c)$ for $c>3^{k'}$}
Next we consider the case when not every component of a message tuple $(\bm{x}_1, \bm{x}_2) \in \classid{1}{\bm{v}} \times  \classid{2}{\bm{w}}$ is present in the largest scalar equivalence class pair. Suppose that $(\classid{1}{\bm{v}}, \classid{2}{\bm{w}})$ is such that
\begin{itemize}
\item at $u_0$ indices from $\mathcal{I}_0(\bm{b}) \cup \mathcal{J}_1(\bm{a}_3)$, the equivalence class pair is either $(\classid{1}{1}, \classid{2}{2})$ or $(\classid{1}{2}, \classid{2}{1})$, and
\item  at $u_1$ indices from $\mathcal{I}_1(\bm{b}) \cup \mathcal{J}_2(\bm{a}_3)$, the equivalence class pair is either $(\classid{1}{1}, \classid{2}{2})$ or $(\classid{1}{2}, \classid{2}{1})$.
\end{itemize}
Let $c > 3^{k'}$ be the index in $\bm{h}_{\bm{b}, \bm{a}_3}$ corresponding to this equivalence class pair. We have that $h_1(1,2) = h_1(2,1)=2$ and $h_2(1,2) = h_2(2,1) = 2$. Then we get that
\begin{equation*}
h_{\bm{b}, \bm{a}_3}^{(c)} = h_{\bm{a}_3}((\bm{v}, \bm{w})_c) = 4^{t_{0,1}- u_0 + t_{1,2} - u_1}\cdot 2^{u_0 + u_1} = \frac{4^{t_{0,1}+t_{1,2}}}{2^{u_0 + u_1}}.
\end{equation*}
Thus the number of components of $\bm{h}_{\bm{b}, \bm{a}_3}$ that have the same value as $h_{\bm{b}, \bm{a}_3}^{(c)}$ are
\begin{equation*}
\binom{t_{0,1}}{u_0} 2^{u_0} \binom{t_{1,2}}{u_1} 2^{u_1} 3^{k'}.
\end{equation*}
Thus, the vector $\bm{h}_{\bm{b}, \bm{a}_3}$ is as follows, where $u_0$ and $u_1$ are indices satisfying $1 \leq u_0 \leq t_{0,1}$ and $1 \leq u_1 \leq t_{1,2}$.
\begin{equation}
\bm{h}_{\bm{b}, \bm{a}_3} = \Big[
\underbrace{4^{t_{0,1} + t_{1,2}} ~\cdots ~ 4^{t_{0,1} + t_{1,2}}}_{3^{k'}} ~~ \cdots ~~
\underbrace{\frac{4^{t_{0,1} + t_{1,2}}}{2^{u_0 + u_1}}  ~\cdots ~ \frac{4^{t_{0,1} + t_{1,2}}}{2^{u_0 + u_1}} }_{\binom{t_{0,1}}{u_0} 2^{u_0} \binom{t_{1,2}}{u_1} 2^{u_1} 3^{k'}} ~~\cdots ~~
\underbrace{\frac{4^{t_{0,1} + t_{1,2}}}{2^{t_{0,1} + t_{1,2}}}  ~\cdots ~ \frac{4^{t_{0,1} + t_{1,2}}}{2^{t_{0,1} + t_{1,2}}} }_{2^{t_{0,1}} 2^{t_{1,2}} 3^{k'}}
\Big]^\top.
\end{equation}
Using Table \ref{tab:illus}, the cardinality of the pre-image set $|A_{123}(\bm{b}, \bm{a}_3)| = 3^{t_{0,0}}8^{t_{0,1}}1^{t_{0,2}}3^{t_{1,0}}1^{t_{1,1}}8^{t_{1,2}}3^{t_{2,0}} = 3^{k'} 8^{t_{0,1}+ t_{1,2}}$. Using this, we can find the value of the entropy as follows.
\begin{align}\label{eq:entropy_low_bd}
&H_{|\mathcal{Z}|}(\bm{h}_{\bm{b}, \bm{a}_3}/|A_{123}(\bm{b}, \bm{a}_3)|) \nonumber \\
&= \sum_{u_0=0}^{t_{0,1}} \sum_{u_1=0}^{t_{1,2}} \binom{t_{0,1}}{u_0}2^{u_0} \binom{t_{1,2}}{u_1}2^{u_1} 3^{k'} \frac{4^{t_{0,1}+t_{1,2}}}{2^{u_0 + u_1}3^{k'}8^{t_{0,1}+ t_{1,2}}} \log_{|\mathcal{Z}|} \Bigg(\frac{2^{u_0 + u_1}3^{k'}8^{t_{0,1}+ t_{1,2}}}{4^{t_{0,1}+t_{1,2}}} \Bigg) \nonumber \\
&= \frac{(t_{0,1}+t_{1,2})\log_{|\mathcal{Z}|}2+k'\log_{|\mathcal{Z}|} 3}{2^{t_{0,1}+t_{1,2}}}  \sum_{u_0=0}^{t_{0,1}} \sum_{u_1=0}^{t_{1,2}} \binom{t_{0,1}}{u_0} \binom{t_{1,2}}{u_1} + \frac{\log_{|\mathcal{Z}|}2}{2^{t_{0,1}+t_{1,2}}} \sum_{u_0=0}^{t_{0,1}} \sum_{u_1=0}^{t_{1,2}} \binom{t_{0,1}}{u_0}u_0 \binom{t_{1,2}}{u_1} \nonumber \\
&\qquad + \frac{\log_{|\mathcal{Z}|}2}{2^{t_{0,1}+t_{1,2}}} \sum_{u_0=0}^{t_{0,1}} \sum_{u_1=0}^{t_{1,2}} \binom{t_{0,1}}{u_0}u_1 \binom{t_{1,2}}{u_1}\nonumber \\
&= k'\log_{|\mathcal{Z}|} 3 + 1.5(t_{0,1}+t_{1,2})\log_{|\mathcal{Z}|} 2 \nonumber \\
&= k\log_{|\mathcal{Z}|} 3 + (1.5\log_{|\mathcal{Z}|}2 - \log_{|\mathcal{Z}|} 3)(t_{0,1}+t_{1,2}) - (t_{0,2} + t_{1,1})\log_{|\mathcal{Z}|} 3.
\end{align}

%
%
%
\end{illustration}
\subsection{Value of $\alpha$}\label{sec:alpha}
Having evaluated $H(\bm{h}_{\bm{b}, \bm{a}_3}/|A_{123}(\bm{b}, \bm{a}_3)|)$, we can find the value of $\alpha$ as used in equation \eqref{eq:sum-rate_lb} in the following manner.
\begin{align}\label{eq:alpha}
  & H(\bm{Z}_1, \bm{Z}_2|f(X_1^k, X_2^k, X_3^k), X_3^k) \nonumber \\
  & = \sum_{\bm{b}} \Pr\{f(X_1^k, X_2^k, X_3^k) = \bm{b}\}\sum_{\bm{x}_3} \Pr\{X_3^k = \bm{x}_3 | f(X_1^k, X_2^k, X_3^k)=\bm{b}\} H(\bm{Z}_1, \bm{Z}_2|f(X_1^k, X_2^k, X_3^k)= \bm{b}, X_3^k=\bm{x}_3) \nonumber\\
  & \geq \sum_{\bm{b}} \Pr\{f(X_1^k, X_2^k, X_3^k) = \bm{b}\}\sum_{\bm{x}_3} \Pr\{X_3^k = \bm{x}_3 | f(X_1^k, X_2^k, X_3^k)=\bm{b}\} H(\bm{h}_{\bm{b}, \bm{a}_3}/|A_{123}(\bm{b}, \bm{a}_3)|) \triangleq \alpha k.
\end{align}
The value of $\Pr\{f(X_1^k, X_2^k, X_3^k) = \bm{b}\}$ can be found from Table \ref{tab:illus} and the i.i.d. uniform assumption on the message tuples. The value of $\Pr\{X_3^k = \bm{x}_3 | f(X_1^k, X_2^k, X_3^k)=\bm{b}\}$ can be evaluated by finding the ratio of the cardinalities of two pre-image sets, i.e., $|A_{123}(\bm{b}, \bm{x}_3)|/|A_{123}(\bm{b})|$. We carry out this computation for the illustrated demand function below.
\begin{illustration}\label{ill:eg_sum-rate}
Consider a realization $\bm{b}$ with $m_1$ 1's, $m_2$ 2's and $k-m_1-m_2$ 0's. Then we have that $\Pr\{f(X_1^k, X_2^k, X_3^k) = \bm{b}\} = (12/27)^{m_1}(3/27)^{m_2}(12/27)^{k-m_1-m_2}$. The number of different demand function realizations which have the same number of 1's, 2's and 0's is $\binom{k}{m_1, m_2, k-m_1-m_2}$. Consider a realization $\bm{a}_3 \in A_3(\bm{b})$ of the message $X_3^k$. The number of message tuples that result in the demand function value $\bm{b}$ and have their $X_3^k$ realization as $\bm{a}_3$ is $|A_{123}(\bm{b}, \bm{a}_3)| =3^{k'}8^{t_{0,1} + t_{1,2}}$, as evaluated in the previous illustration. The number of message tuples in the pre-image set $A_{123}(\bm{b})$ can be evaluated using Table \ref{tab:illus} as $|A_{123}(\bm{b})| = 12^{m_1}3^{m_2}12^{k - m_1 - m_2} = 3^k4^{k-m_2}$. Because of the uniform i.i.d. assumption, we have that
\begin{align*}
  \Pr\{X_3^k = \bm{a}_3 | f(X_1^k, X_2^k, X_3^k) = \bm{b}\} & = \frac{|A_{123}(\bm{b}, \bm{a}_3)|}{|A_{123}(\bm{b})|} = \frac{3^{k'}8^{t_{0,1 + t_{1,2}}}}{3^k4^{k-m_2}}\\
   & = (1/4)^{k-m_1-m_2-t_{0,1}-t_{0,2}} (2/3)^{t_{0,1}} (1/12)^{t_{0,2}} (1/4)^{m_1-t_{1,1}-t_{1,2}} (2/3)^{t_{1,2}} (1/12)^{t_{1,1}}.
\end{align*}
Using equations \eqref{eq:alpha}, \eqref{eq:entropy_low_bd} and the probabilities computed above, 
the value of $\alpha$ is found in Appendix \ref{app:eg_alpha_val} to be
\begin{equation}\label{eq:eg_alpha_val}
  \alpha = \frac{8}{9}\log_{|\mathcal{Z}|}2+\frac{4}{12}\log_{|\mathcal{Z}|} 3.
\end{equation}
Using this value of $\alpha$ in equation \eqref{eq:sum-rate_lb}, we get that
\begin{align*}
\frac{R_1 + R_2}{2} + \epsilon \geq \frac{\alpha + \big(\log_{|\mathcal{Z}|} 9 -\frac{8}{9} \log_{|\mathcal{Z}|}4\big) }{2\log_{|\mathcal{Z}|} |\mathcal{A}|} = \frac{\frac{8}{9}\log_{|\mathcal{A}|}2 + \frac{1}{3}\log_{|\mathcal{A}|}3 + 2\log_{|\mathcal{A}|}3 - \frac{16}{9}\log_{|\mathcal{A}|}2}{2}\approx \frac{1.7725}{2}.
\end{align*}

\end{illustration}

\subsection{Example demand function: sum over $GF(2)$}\label{subsec:GF2_sum}
Suppose the messages $X_1, X_2, X_3 \in \{0,1\}$ and the terminal wants to compute their finite field sum over $GF(2)$. For this demand function, we demonstrate that the outer bound to the rate region is tight. We assume that the alphabet used for communication is $\mathcal{Z} = \{0, 1\}$. 
For $X_3$ we have that $0 \not \equiv 1$, and thus for a function $g(X_3^k)$ that returns the equivalence class that $X_3^k$ belongs to, we have that $H(g(X_3^k)) = k$. Thus from Lemma \ref{lem:X_3eq} we get that $R_{31} + R_{32} \geq 1$.

For $X_u,\: u =1$ or $2$, using the values of the finite field sum, we obtain the following partitions
\begin{equation*}
\overset{0}{\equiv}|_u: \{0\} \cup \{1\} \:\text{ and }\: \overset{1}{\equiv}|_u : \{0\} \cup \{1\}.
\end{equation*}
Thus, 
for a $GF(2)$-sum realization $\bm{b}$ and $\bm{a}_3 \in A_3(\bm{b})$, the equivalence classes satisfy $|\eqvclass{1}{\bm{a}_3}{\bm{v}}| = |\eqvclass{2}{\bm{x}_3}{\bm{w}}|=1$ for any class index $\bm{v}$ or $\bm{w}$ and $V_1(\bm{a}_3) = V_2(\bm{a}_3) = 2^k$.
Thus the vector $\bm{d}_u(\bm{a}_3)$ defined in Lemma \ref{lem:Z_u-majorize} is $\bm{1}_{2^k}$, and hence $H(\bm{d}_u(\bm{a}_3)/2^k) = k$, giving us the value of $\gamma$ as
\begin{equation*}
\gamma = 1 \implies R_u + \epsilon \geq 1.
\end{equation*}
For any value of $X_1^k, X_3^k$ and $f(X_1^k,X_2^k,X_3^k)$ the value of $X_2^k$ is fixed by
$X_2^k = f(X_1^k,X_2^k,X_3^k) - X_1^k - X_3^k$, where the subtraction operations are also over $GF(2)$. Hence we have that
\begin{equation*}
  h_{\bm{a}_{3}}(\bm{v}, \bm{w}) = 1 \text{ for every } \bm{a}_{3} \in A_3(\bm{b}) ~\text{ and every }~ (\classid{1}{\bm{v}},\classid{2}{\bm{w}}) \in \mathcal{V}_{12}(\bm{a}_{3}, \bm{b}).
\end{equation*}
We enumerate the different $(X_1, X_2)$ pairs that result in $b^{(i)}$ for the realization $a_3^{(i)}$ in different cases as below.
\begin{itemize}
\item If $i \in \mathcal{I}_0(\bm{b}) \cap \mathcal{J}_0(\bm{a}_3)$: $(X_1, X_2) \in \{(0, 0), (1, 1)\}$.
\item If $i \in \mathcal{I}_0(\bm{b}) \cap \mathcal{J}_1(\bm{a}_3)$: $(X_1, X_2) \in \{(0, 1), (1, 0)\}$.
\item If $i \in \mathcal{I}_1(\bm{b}) \cap \mathcal{J}_0(\bm{a}_3)$: $(X_1, X_2) \in \{(0, 1), (1, 0)\}$.
\item If $i \in \mathcal{I}_1(\bm{b}) \cap \mathcal{J}_1(\bm{a}_3)$: $(X_1, X_2) \in \{(0, 0), (1, 1)\}$.
\end{itemize}
Since there are two choices in each case, we have that $|A_{123}(\bm{b}, \bm{a}_3)| = 2^k$. Thus we have that $\bm{h}_{\bm{b}, \bm{a}_3}/|A_{123}(\bm{b}, \bm{a}_3)|=\bm{1}_{2^k}/2^k$ which gives the value of $\alpha$ as
\begin{equation*}
\alpha = 1 \implies (R_1 + R_2)/2 + \epsilon \geq (1 + H(f(X_1^k, X_2^k, X_3^k))/k)/2 = (1 + k/k)/2 = 1.
\end{equation*}
We describe simple network code that allows $t$ to compute the $GF(2)$-sum by carrying out the componentwise addition $\bm{Z}_{1}(X_1^k, \bm{Z}_{31}(X_3^k)) + \bm{Z}_2(X_2^k, \bm{Z}_{32}(X_3^k))$. The codewords used, for any $c \in \mathbb{Z}_{\geq 0}$ such that $c \leq k$, are
\begin{align*}
  \bm{Z}_{31}(X_3^k) &= (X_3^{(1)}, X_3^{(2)}, \ldots, X_3^{(c)}),\:\: \bm{Z}_{32}(X_3^k) = (X_3^{(c+1)}, X_3^{(c+2)}, \ldots, X_3^{(k)}),\\
  \bm{Z}_1(X_1^k, \bm{Z}_{31}(X_3^k)) &= (X_1^{(1)}+X_3^{(1)}, \ldots, X_1^{(c)}+X_3^{(c)}, X_1^{(c+1)}, \ldots, X_1^{(k)}),\\
  \bm{Z}_2(X_2^k, \bm{Z}_{32}(X_3^k)) &= (X_2^{(1)},\ldots, X_2^{(c)}, X_2^{(c+1)}+X_3^{(c+1)}, \ldots, X_2^{(k)}+X_3^{(k)}).
\end{align*}
All operations above are over $GF(2)$.
Then $\mathbb{E}\ell (\bm{Z}_{31}) = c, \mathbb{E}\ell (\bm{Z}_{32}) = k-c$. 
If $X_1, X_3 \sim \text{Unif}\{0, 1\}$ then $X_1 + X_3 \sim \text{Unif}\{0, 1\}$ and similarly for $X_2 + X_3$. Hence $\mathbb{E}\ell (\bm{Z}_1) = \mathbb{E}\ell (\bm{Z}_2) = k$. These imply that the rate tuple achieved is
\begin{equation*}
(R_{31}, R_{32}, R_1, R_2) = (c/k, 1-c/k, 1, 1),
\end{equation*}
which matches the lower bounds derived above.

\subsection{Example demand function: arithmetic sum}\label{subsec:arith_sum}
Suppose the message alphabet is $\mathcal{A}=\{0,1\}$, such that the messages $X_1,X_2,X_3$ are independent bits each equally likely to be $0$ or $1$. The demand function $f(X_1,X_2,X_3)=X_1+X_2+X_3$ is the sum of the messages over the integers, such that $\mathcal{B}=\{0,1,2,3\}$. We use the codeword alphabet $\mathcal{Z}=\{0,1\}$. This case of arithmetic sum computation in the variable-length network code framework was considered in \cite{tripathyR16} and we recover the results there in our general framework.
For a given value of the arithmetic sum $\bm{b} \in \{0,1,2,3\}^k$ the set $A_3(\bm{b})$ of valid realizations for $X_3^k$ can be described as
\begin{equation*}
  A_3(\bm{b}) = \{\bm{a}_3 \in \{0,1\}^k: a_3^{(i)}=0 \text{ if } \bm{b}^{(i)}=0 \text{ and } a_3^{(j)}=1 \text{ if } \bm{b}^{(j)}=3 \text{ for all } i,j \in \{1,2,\ldots, k\}\}.
\end{equation*}
We consider the equivalence relation $\overset{\bm{a}_3}{\equiv}|_{1}$ for the arithmetic sum demand function. Then
\begin{equation*}
  \bm{x}_1 \overset{\bm{a}_3}{\equiv} \bm{y}_1|_{1} \Leftrightarrow \bm{x}_1 = \bm{y}_1,
\end{equation*} because if $\bm{x}_1 \neq \bm{y}_1$ then for all $\bm{x}_2 \in \{0,1\}^k$ we have that $\bm{x}_1 + \bm{x}_2 + \bm{a}_3 \neq \bm{y}_1 + \bm{x}_2 + \bm{a}_3$. A similar conclusion also holds true for the $\overset{\bm{a}_3}{\equiv}|_{2}$ relation. Thus for both $u=1$ and $2$, we have that $|\eqvclass{u}{\bm{a}_3}{\bm{v}}|=1$ for any class index $\bm{v}$ and $V_u(\bm{a}_3) = 2^k$.
We use the alphabet $\mathcal{Z}=\{0,1\}$ for communication. The vector $\bm{d}_u(\bm{a}_3)$ defined in Lemma \ref{lem:Z_u-majorize} satisfies $\bm{d}_u(\bm{a}_3) = \bm{1}_{2^k}$ in this case, where $\bm{1}_t$ indicates a vector of ones with length $t$. Using this in equation \eqref{eq:gamma}, we obtain the value of $\gamma$ as
\begin{equation*}
  \gamma = \frac{1}{k} \sum_{\bm{a}_3 \in \mathcal{A}^k} \Pr\{X_3^k = \bm{a}_3\}H_{|\mathcal{Z}|} \Big(\frac{\bm{d}_u(\bm{a}_3)}{|\mathcal{A}|^k}\Big) = \frac{1}{k}\cdot k = 1,
\end{equation*} and thus $R_u + \epsilon \geq 1 / \log_{|\mathcal{Z}|}|\mathcal{A}| = 1$.

For any value of $X_1^k, X_3^k$ and $f(X_1^k,X_2^k,X_3^k)$ the value of $X_2^k$ is fixed by
$X_2^k = f(X_1^k,X_2^k,X_3^k) - X_1^k - X_3^k$.
Hence, for every $(\classid{1}{\bm{v}},\classid{2}{\bm{w}})\in \mathcal{V}_{12}(\bm{a}_3, \bm{b})$ there is exactly one message tuple whose $X_1^k$ and $X_2^k$ belong to the equivalence classes $\eqvclass{1}{\bm{a}_3}{\bm{v}}$ and $\eqvclass{2}{\bm{a}_3}{\bm{w}}$ respectively. Thus we have that
\begin{equation*}
  h_{\bm{a}_{3}}(\bm{v}, \bm{w}) = 1 \text{ for every } \bm{a}_{3} \in A_3(\bm{b}) ~\text{ and every }~ (\classid{1}{\bm{v}},\classid{2}{\bm{w}}) \in \mathcal{V}_{12}(\bm{a}_{3}, \bm{b}).
\end{equation*}
Consider an arithmetic sum realization $\bm{b}$ with $m_0$ 0's, $m_1$ 1's, $m_2$ 2's and $k - m_0 - m_1 - m_2$ 3's.
Define $t_{p,q} \triangleq |\mathcal{I}_p(\bm{b}) \cap \mathcal{J}_q(\bm{a}_3)|$ for every $p \in \{0,1,2,3\}$ and $q \in \{0,1\}$. Then for any choice of $\bm{b}$ and $\bm{a}_3 \in A_3(\bm{b})$, $t_{0,1} = t_{3,0} = 0$. The cardinality of the pre-image set $|A_{123}(\bm{b}, \bm{a}_3)| = 2^{t_{1,0}+t_{2,1}} =2^{t_{1,0}+m_2-t_{2,0}}$. The value of the entropy $H(\bm{h}_{\bm{b}, \bm{a}_3}/|A_{123}(\bm{b}, \bm{a}_3)|) = t_{1,0} + m_2 - t_{2,0}$. From the function definition, we can check that $|A_{123}(\bm{b})| = 3^{m_1+m_2}$. Thus we have that
\begin{equation*}
\Pr\{X_3^k = \bm{x}_3|f(X_1^k, X_2^k, X_3^k) = \bm{b}\} = \frac{|A_{123}(\bm{b}, \bm{a}_3)|}{|A_{123}(\bm{b})|} = \frac{2^{t_{1,0}+m_2-t_{2,0}}}{3^{m_1+m_2}}.
\end{equation*}
Then the value of $\alpha$ can be found as follows.
\begin{align*}
\alpha k&= \sum_{\bm{b}} \Pr\{f(X_1^k, X_2^k, X_3^k) = \bm{b}\}\sum_{\bm{a}_3 \in A_3(\bm{b})} \Pr\{X_3^k = \bm{a}_3 | f(X_1^k, X_2^k, X_3^k)=\bm{b}\} H(\bm{h}_{\bm{b}, \bm{a}_3} / A_{123}(\bm{b}, \bm{a}_3))\\
&= \sum_{\bm{b}} \Pr\{f(X_1^k, X_2^k, X_3^k) = \bm{b}\} \sum_{t_{1,0} = 0}^{m_1}\sum_{t_{2,0} = 0}^{m_2}  \frac{m_1!(2/3)^{t_{1,0}}(1/3)^{m_1-t_{1,0}}}{t_{1,0}!(m_1-t_{1,0})!} \frac{m_2!(2/3)^{m_2-t_{2,0}}(1/3)^{t_{2,0}}}{t_{2,0}!(m_2-t_{2,0})!} (t_{1,0}+m_2-t_{2,0})\\
&= \sum_{m_1=0}^{k}\sum_{m_2=0}^{k-m_1} \frac{k!2^{k-m_1-m_2}}{m_1!m_2!(k-m_1-m_2)!}\left(\frac{1}{8}\right)^{k-m_1-m_2}\left(\frac{3}{8} \right)^{m_1+m_2} \frac{2(m_1+m_2)}{3}\\
&=\frac{2/3}{4^k}\sum_{m_1=0}^{k}\sum_{m_2=0}^{k-m_1} \frac{k!(m_1+m_2)}{m_1!m_2!(k-m_1-m_2)!} \bigg(\frac{3}{2}\bigg)^{m_1+m_2}
= \frac{2/3}{4^k} 2\cdot \frac{3}{2} k4^{k-1} = 0.5k.
\end{align*}
Putting this value of $\alpha$ in equation \eqref{eq:sum-rate_lb}, we get that
\begin{equation*}
\frac{R_1 + R_2}{2} + \epsilon \geq \frac{0.5 + 3 - 0.75\log 3}{2} \approx \frac{2.31128}{2}.
\end{equation*}
\begin{remark}
We note that the lower bound for the sum rate shown above is tighter than the bound $R_1 + R_2 > 2.25$ obtained in \cite{tripathyR16} for the same problem. The inequalities considered there were similar to those used in arriving at the sum rate lower bound in equation \eqref{eq:sum-rate_lb}, however, in \cite{tripathyR16} they did not include $X_3^k$ in the conditioning while lower bounding $H(\bm{Z}_1, \bm{Z}_2|f(X_1^k, X_2^k, X_3^k))$ as done here. Instead, they directly lower bounded $H(\bm{Z}_1, \bm{Z}_2|f(X_1^k, X_2^k, X_3^k))$ using a so-called \textit{clumpy} distribution, which is a p.m.f. that majorizes any valid conditional p.m.f. of the pair $(\bm{Z}_1, \bm{Z}_2)$ given $f(X_1^k, X_2^k, X_3^k)$. Since only a lower bound to the entropy of the clumpy distribution was obtained in \cite{tripathyR16}, the corresponding lower bound for the sum rate ends up being looser than the value we obtain here.
\end{remark}

\section{Conclusions and future work}\label{sec:conclude_future}
In this paper, we have described a procedure to obtain an outer bound for the rate region (in the setup of \cite{songYC03}) for computing a function with zero-error over a simple DAG network. The demand function can be an arbitrary discrete-valued function and only needs to be specified as a function table. For computing the arithmetic sum, we show that the outer bound obtained is tighter than the one in \cite{tripathyR16}. For computing the $GF(2)$-sum, we show that the lower bounds for the rate tuple obtained using our procedure can also be achieved by a simple network code.
Our method uses the equivalence relations defined in \cite{guangYL16} as adapted to the specific DAG network considered here. Assuming a independent, uniform probability distribution for each of the messages, we compute the probability that the messages belong to a particular equivalence class. These are used in obtaining a lower bound to the conditional entropy of the descriptions transmitted on the edges, which imply an outer bound to the rate region.

There are several opportunities for future work. Our investigation reveals that the analysis of even very simple, non-tree networks require the usage of fairly nontrivial techniques. Nevertheless, it is evident that taking into account the compressibility of the sources allows for higher computation rates. It would be interesting to consider the application of our techniques (or simplifications thereof) to more complicated topologies. It may also be fruitful to consider specific functions of interest for which the equivalence classes (used extensively in Section \ref{sec:outer_bd}) have a simpler characterization.

\ifCLASSOPTIONcaptionsoff
  \newpage
\fi



\bibliographystyle{IEEEtran}
%

\appendices
\section{Proof of Lemma \ref{lem:best-NS}}\label{app:best-NS}
\begin{IEEEproof}
Let $l_1\leq l_2\leq \cdots \leq l_{\max}$ be the lengths over $\mathcal{Z}$ of the best non-singular code for $\bm{Z}$. We demonstrate a function $g$ such that the set of lengths $\{g(l_i): i = 1,2,\ldots\}$ satisfy Kraft's inequality, i.e., $\sum_{i=1}^{\infty}|\mathcal{Z}|^{-g(l_i)} \leq 1$. Choose $g(l_i) \triangleq l_i + 2\lfloor \log_{|\mathcal{Z}|}(l_i+ |\mathcal{Z}|-1) \rfloor$. Since there are $|\mathcal{Z}|^{l_i}$ different non-singular codewords with length $l_i$ and if $l_{\max} > l_i$ then all of them must have been used in the best non-singular code, we have that
\begin{align*}
\sum_{i} |\mathcal{Z}|^{-l_i-2 \lfloor\log_{|\mathcal{Z}|}(l_i+|\mathcal{Z}|-1)\rfloor} &= \sum_{l_1}^{l_{\max}} |\mathcal{Z}|^{l_i}|\mathcal{Z}|^{-l_i-2 \lfloor \log_{|\mathcal{Z}|}(l_i+|\mathcal{Z}|-1)\rfloor}
\leq \sum_{l=1}^{\infty} |\mathcal{Z}|^{-(2\lfloor \log_{|\mathcal{Z}|}(l+|\mathcal{Z}|-1)\rfloor)}\\
&= \sum_{l=1}^{|\mathcal{Z}|^2-|\mathcal{Z}|} \frac{1}{|\mathcal{Z}|^2} + \sum_{l=|\mathcal{Z}|^2 - |\mathcal{Z}| + 1}^{|\mathcal{Z}|^3-|\mathcal{Z}|} \frac{1}{|\mathcal{Z}|^4} + \sum_{l=|\mathcal{Z}|^3-|\mathcal{Z}|+1}^{|\mathcal{Z}|^4-|\mathcal{Z}|} \frac{1}{|\mathcal{Z}|^6} + \cdots \\
&= \frac{|\mathcal{Z}|-1}{|\mathcal{Z}|} + \frac{|\mathcal{Z}|-1}{|\mathcal{Z}|^2} + \frac{|\mathcal{Z}|-1}{|\mathcal{Z}|^3} + \cdots = 1.
\end{align*}
The above is also true if $l_{\max} \rightarrow \infty$. Thus there exists an uniquely decodable code for $\bm{Z}$ whose codeword lengths are $\{\lceil g(l_i)\rceil :i = 1,2,\ldots\}$. Using random variable $\ell$ to denote the lengths of the codewords in the best non-singular code for $\bm{Z}$, we have that
\begin{align*}
H_{|\mathcal{Z}|}(\bm{Z}) &\leq \E(\ell + 2\lfloor \log_{|\mathcal{Z}|}(\ell+|\mathcal{Z}|-1) \rfloor) \leq \E \ell + 2\E \log_{|\mathcal{Z}|}(\ell+|\mathcal{Z}|-1) \overset{\text{(i)}}{\leq} \E \ell + 2\log_{|\mathcal{Z}|}(\E \ell + |\mathcal{Z}| - 1)\\
&\leq \E \ell + 2\log_{|\mathcal{Z}|}(1+H_{|\mathcal{Z}|} + |\mathcal{Z}| - 1) \implies \E \ell \geq H_{|\mathcal{Z}|}(\bm{Z}) -2\log_{|\mathcal{Z}|}(|\mathcal{Z}|+ H_{|\mathcal{Z}|}(\bm{Z})),
\end{align*}
where inequality (i) above is true due to Jensen's inequality.
\end{IEEEproof}

\section{Proof of Lemma \ref{lem:distinct_Z1_Z2_label}}\label{app:proof_distinct}
\begin{IEEEproof}
We use the same argument for $u=1$ and $2$. Partition the set $\mathcal{A}^k$ into $\prod_{i=1}^{k}V_{u}(a_3^{(i)})$ 
disjoint subsets based on the equivalence relation
$\overset{a_3^{(i)}}{\equiv}|_{u}$ 
in each component $i \in \{1,2,\ldots,k\}$. Every element $(x_u^{(1)},\allowbreak x_u^{(2)},\allowbreak \ldots,\allowbreak x_u^{(k)}) \in \mathcal{A}^k$ belongs to a subset in the partition. Suppose the number of distinct $\bm{Z}_u$-labels transmitted on $(s_u,t)$ is strictly less than $\prod_{i=1}^{k}V_{u}(a_3^{(i)})$. 
Then by the pigeon-hole principle, there exist two elements $\bm{x}_u \triangleq (x_u^{(1)},\allowbreak x_u^{(2)},\allowbreak \ldots,\allowbreak x_u^{(k)})$ and $\bm{y}_u \triangleq (y_u^{(1)},\allowbreak y_u^{(2)},\allowbreak \ldots,\allowbreak y_u^{(k)})$ that belong to different equivalence relation subsets of $\mathcal{A}^k$ but are given the same $\bm{Z}_u$-label. 
Let $J \subseteq \{1,2,\ldots,k\}$ be the index set collecting all indices $j$ such that $x_u^{(j)} \overset{a_3^{(j)}}{\not\equiv} y_u^{(j)}|_{u}$ 
Since $\bm{x}_u$ and $\bm{y}_u$ belong to different equivalence relation partitions, $J \neq \emptyset$ and for every $j \in J$ there exists, by definition, a $a_v^{(j)} \in \mathcal{A}, v \in \{1,2\}\setminus u$ such that
\begin{equation*}
  f\left(x_u^{(j)},a_v^{(j)},a_3^{(j)}\right) \neq f\left(y_u^{(j)},a_v^{(j)},a_3^{(j)}\right).
\end{equation*}
Then consider the following two different scenarios of $k$ independent message realizations.
\begin{IEEEeqnarray*}{uL}
(I)\: & X_u^k= \bm{x}_u , X_v^k = \bm{x}_v, X_3^k= \bm{a}_3 \\
(II)\: & X_u^k= \bm{y}_u, X_v^k = \bm{x}_v , X_3^k=\bm{a}_3,
\end{IEEEeqnarray*}
where $\bm{x}_v$ is such that $x_v^{(j)} = a_v^{(j)}$ for every $j \in J$.
Note that the realizations of $X_v^k$ and $X_3^k$ are the same in both cases, and hence so is the label transmitted on the $(s_v,t)$ edge. On the other hand, by assumption we have that the label transmitted on the edge $(s_u,t)$ is the same in both cases as well. Then the terminal cannot recover the correct value of the demand function for the components in the set $J$, as the $(\bm{Z}_1, \bm{Z}_2)$-labels received are the same but the function values at the components in $J$ are different by choice of $\bm{x}_u, \bm{y}_u$. This contradicts the fact that the network code allows $t$ to recover $f(X_1^k,X_2^k,X_3^k)$ with zero error.
The two scenarios considered above have $\bm{x}_u \overset{\bm{a}_3}{\not \equiv}\bm{y}_u |_u$ and thus also give a proof for the second statement of the lemma.
\end{IEEEproof}

\section{Calculation for equation \eqref{eq:eg_alpha_val}}\label{app:eg_alpha_val}
Using equations \eqref{eq:alpha}, \eqref{eq:entropy_low_bd} and the probabilities computed in illustration \ref{ill:eg_sum-rate}, we can find the value of $\alpha$ as follows.
\begin{align*}
  & \alpha k  \\
  & =\! \sum_{\bm{b}} \Pr\{f(X_1^k, X_2^k, X_3^k) = \bm{b}\}\!\!\! \sum_{\bm{x}_3 \in A_3(\bm{b})}\!\!\! \Pr\{X_3^k = \bm{x}_3 | f(X_1^k, X_2^k, X_3^k)\!=\!\bm{b}\} \Big(k\log_{|\mathcal{Z}|} 3 - (t_{0,2} + t_{1,1})\log_{|\mathcal{Z}|} 3\\
  & \pushright{+ (1.5\log_{|\mathcal{Z}|}2 - \log_{|\mathcal{Z}|} 3)(t_{0,1}+t_{1,2})\Big)}\\
  & = \sum_{\bm{b}} \Pr\{f(X_1^k, X_2^k, X_3^k) = \bm{b}\} \Bigg[k\log_{|\mathcal{Z}|} 3\\
  & + \sum_{t_{1,1}=0}^{m_1}\sum_{t_{1,2}=0}^{m_1-t_{1,1}}\binom{m_1}{t_{1,1}, t_{1,2}, m_1-t_{1,1}-t_{1,2}} (\frac{1}{4})^{m_1-t_{1,1}-t_{1,2}} (\frac{2}{3})^{t_{1,2}} (\frac{1}{12})^{t_{1,1}} ((\log_{|\mathcal{Z}|}\frac{2^{1.5}}{3}) t_{1,2}-t_{1,1}\log_{|\mathcal{Z}|} 3)\\
  &~~\cdot \sum_{t_{0,1}=0}^{k-m_1-m_2} \sum_{t_{0,2}=0}^{k-m_1-m_2-t_{0,1}} \binom{k-m_1-m_2}{t_{0,1}, t_{0,2}, k-m_1-m_2-t_{0,1}-t_{0,2}} (\frac{1}{4})^{k-m_1-m_2-t_{0,1}-t_{0,2}} (\frac{2}{3})^{t_{0,1}} (\frac{1}{12})^{t_{0,2}} \\
  & + \sum_{t_{0,1}=0}^{k-m_1-m_2}\sum_{t_{0,2}=0}^{k-m_1-m_2-t_{0,1}}\binom{k-m_1-m_2}{t_{0,1}, t_{0,2}, m_1-t_{0,1}-t_{0,2}} (\frac{1}{4})^{k-m_1-m_2-t_{0,1}-t_{0,2}} (\frac{2}{3})^{t_{0,1}} (\frac{1}{12})^{t_{0,2}}\\
 &\pushright{\cdot ((\log_{|\mathcal{Z}|}\frac{2^{1.5}}{3})t_{0,1}-t_{0,2}\log_{|\mathcal{Z}|} 3)} \\
  &\pushright{\cdot \sum_{t_{1,1}=0}^{m_1} \sum_{t_{1,2}=0}^{m_1-t_{1,1}} \tbinom{m_1}{t_{1,1}, t_{1,2}, m_1-t_{1,1}-t_{1,2}} (\frac{1}{4})^{m_1-t_{1,1}-t_{1,2}} (\frac{2}{3})^{t_{1,2}} (\frac{1}{12})^{t_{1,1}}\Bigg]}\\
  & = \sum_{\bm{b}} \Pr\{f(X_1^k, X_2^k, X_3^k) = \bm{b}\} \Bigg[k\bigg(\log_{|\mathcal{Z}|}2+\frac{\log_{|\mathcal{Z}|} 3}{4}\bigg) + m_2\bigg(-\log_{|\mathcal{Z}|}2 + \frac{9}{12}\log 3 \bigg)\Bigg]\\
  & = \sum_{m_1 = 0}^{k} \sum_{m_2 = 0}^{k-m_1} \binom{k}{m_1, m_2, k-m_1-m_2} \big(\frac{4}{9}\big)^{k-m_1-m_2} \big(\frac{4}{9}\big)^{m_1} \big(\frac{1}{9}\big)^{m_2} \Bigg[k\bigg(\log_{|\mathcal{Z}|}2+\frac{\log_{|\mathcal{Z}|} 3}{4}\bigg)\\
  &\pushright{+ m_2\bigg(-\log_{|\mathcal{Z}|}2 + \frac{9}{12}\log_{|\mathcal{Z}|} 3 \bigg)\Bigg]}\\
  & = k\bigg(\log_{|\mathcal{Z}|}2+\frac{\log_{|\mathcal{Z}|} 3}{4}\bigg) \\
  & \quad + \bigg(-\log_{|\mathcal{Z}|}2 + \frac{9}{12}\log_{|\mathcal{Z}|} 3 \bigg) \sum_{m_1 = 0}^{k} \sum_{m_2 = 0}^{k-m_1} \binom{k}{m_1, m_2, k-m_1-m_2} m_2 \big(\frac{1}{9}\big)^{m_2} \big(\frac{4}{9}\big)^{k-m_2}\\
  & = k\bigg(\log_{|\mathcal{Z}|}2+\frac{\log_{|\mathcal{Z}|} 3}{4}\bigg) +\bigg(-\log_{|\mathcal{Z}|}2 + \frac{9}{12}\log_{|\mathcal{Z}|} 3 \bigg) \frac{k}{9}.
\end{align*}
\end{document}